\documentclass[a4paper,11pt]{article}

\pdfoutput=1
\usepackage{comment}
\usepackage{fixltx2e}
\usepackage[table,xcdraw]{xcolor}
\usepackage{jcappub} 
\usepackage{float} 
\usepackage[scr=boondox]{mathalfa}
\usepackage{bm}
\usepackage{stmaryrd}
\usepackage{bbold}
\usepackage[normalem]{ulem}
\usepackage{amsmath}
\usepackage{bbm}
\usepackage{graphbox}
\usepackage{subfig}
\usepackage{booktabs}
\usepackage{hyperref}

\def\be{\begin{equation}}
\def\ee{\end{equation}}
\def\bea{\begin{eqnarray}}
\def\eea{\end{eqnarray}}
\newcommand{\vs}{\nonumber\\}
\def\ba#1\ea{\begin{align}#1\end{align}}
\def\bg#1\eg{\begin{gather}#1\end{gather}}

\def\Mpch{\,h^{-1}\,{\rm Mpc}}
\def\iMpch{\,h\,{\rm Mpc}^{-1}}

\def\leftfield{\texttt{LEFTfield}}

\newcommand{\s}{\sigma}

\newcommand{\refeq}[1]{Eq.~(\ref{eq:#1})}          
\newcommand{\refeqs}[2]{Eqs.~(\ref{eq:#1})--(\ref{eq:#2})}          
          
\newcommand{\reffig}[1]{Fig.~\ref{fig:#1}}

\newcommand{\refsec}[1]{Sec.~\ref{sec:#1}}          
\newcommand{\refapp}[1]{App.~\ref{app:#1}}

\newcommand{\unit}{{\mathbb{1}}}

\def\Plin{P_{\rm L}}

\renewcommand{\v}[1]{\bm{#1}}

\renewcommand{\emph}[1]{\textit{#1}}

\newcommand{\vx}{\v{x}}
\newcommand{\vy}{\v{y}}

\newcommand{\vk}{\v{k}}

\newcommand{\vp}{\v{p}}

\def\Del{\mathcal{D}}

\def\dgdet{\delta_{g,\rm det}}

\def\dg{\delta_{g}}
\def\kmax{k_\text{max}}

\def\dgkmax{\delta_{g,\kmax}}

\def\hdg{\hat{\delta}_{g}}
\def\hdgkmax{\hat{\delta}_{g,\kmax}}

\newcommand{\<}{\langle}
\renewcommand{\>}{\rangle}

\DeclareMathOperator{\ii}{i}

\renewcommand{\d}{\delta}

\newcommand{\K}{\mathcal{K}}
\newcommand{\C}{\mathcal{C}}
\newcommand{\Csym}{\mathcal{C}_{\rm symm}}
\newcommand{\bb}{\mathcal{B}}

\def\Mpch{\,h^{-1}\text{Mpc}}
\def\iMpch{\,h\,\text{Mpc}^{-1}}

\def\Plin{P_\text{L}}

\def\P{\mathcal{P}}

\def\N{\mathcal{N}}

\def\L{\Lambda}

\def\LL{\mathcal{L}}
\def\hatP{\mathcal{P}^\prime} 
\def\vhatSigma{\hat{\v{\Sigma}}}
\def\hatSigma{\hat{\Sigma}}

\def\vKtwoinverse{[\bm{\mathcal{K}}^{\{2\}}]^{-1}}

\def\ker{K}

\def\K{\mathcal{K}}

\def\CO#1{C^{\{#1\}}}
\def\COreal#1{C^{{\rm r,}\{#1\}}}

\def\beps#1{b^{\{#1\}}}

\def\epsG{\eps_{\rm G}}

\newcommand{\eps}{\epsilon}
\newcommand{\epsold}{\epsilon^{\rm n-min}}
\newcommand{\epslin}{\epsilon_{m=1}}

\def\din{\delta_{\rm in}}
\def\dinL{\delta_{\rm in,\Lambda}}

\def\otd{\text{td}}
\def\Otd{O_{\otd}}

\newcommand{\perm}[1]{ \expandafter\ifstrempty\expandafter{#1} {\mbox{perm.}} {\mbox{$#1$ perm.}} }

\makeatletter
\newlength{\apb@width}
\newcommand{\autoparbox}[2][c]{\settowidth{\apb@width}{#2}\parbox[#1]{\apb@width}{#2}}

\makeatother

\def\lapl{\nabla^2}
\def\dirac#1{\delta_{\mathrm{D}}^{[#1]}}
\def\diracc{\delta_{\rm D}}
\def\diracpi{\hat{\delta}_{\rm D}}
\def\diracindex#1{\hat{\delta}_{\mathrm{D},#1}}

\renewcommand{\comment}[1]{}

\title{Non-Gaussian Galaxy Stochasticity and the Noise-Field Formulation}

\author[a,b,c]{Henrique Rubira,}
\author[d]{Fabian Schmidt}

\affiliation[a]{University Observatory, Faculty of Physics, Ludwig-Maximilians-Universit\"at, Scheinerstr. 1, D-81679 München, Germany}
\affiliation[b]{Kavli Institute for Cosmology Cambridge, Madingley Road, Cambridge CB3 0HA, UK}
\affiliation[c]{Centre for Theoretical Cosmology, Department of Applied Mathematics and Theoretical Physics
	University of Cambridge, Wilberforce Road, Cambridge, CB3 0WA, UK}
    
\affiliation[d]{Max-Planck-Institut f\"{u}r Astrophysik,\\ 
Karl-Schwarzschild-Str. 1, 85748 Garching, Germany}

\emailAdd{fabians@mpa-garching.mpg.de}
\emailAdd{henrique.rubira@lmu.de}

\abstract{
  We revisit the stochastic, or noise, contributions to the galaxy density field within the effective field theory (EFT) of large-scale structure. Starting from the general, all-order expression of the EFT partition function, we elucidate how the stochastic contributions can be described
  by local nonlinear couplings of a \emph{single Gaussian noise field}. We introduce an alternative formulation of the partition function in terms of such a noise field, and derive the corresponding field-level likelihood for biased tracers. This noise-field formulation can capture the complete set of stochastic contributions to the galaxy density at the field level in a normalized, positive-definite probability density which is suitable for numerical sampling. We illustrate this by presenting the first results of EFT-based field-level inference with non-Gaussian and density-dependent stochasticity on dark matter halos using \texttt{LEFTfield}.
}


\begin{document}

\maketitle
\flushbottom

\section{Introduction}\label{sec:intro}

Surveys that probe tracers of matter are the backbone of modern cosmology. In this context, the general bias expansion (see \cite{Desjacques:2016bnm} for a review) offers a symmetry-based framework to connect observed tracers to the underlying dark matter density field. While in general nonlocal in time, this bias expansion can be written as a local-in-time relation either in the initial conditions (Lagrangian approach) or the evolved density field (Eulerian approach), thanks to the factorization of time- and space-dependences in perturbation theory, provided linear growth is scale-independent.

Many investigations of the statistics of biased tracers (often simply referred to as ``galaxies'' in the following) in the EFT have focused on the deterministic part, i.e. the bias expansion,
\be
\dgdet(\vx,\tau) = \sum_O b_O(\tau) O(\vx,\tau)\,,
\label{eq:dgdetL_intro}
\ee
which describes the galaxy density in the mean-field sense. The operators $O(\vx,\tau)$ encode the dependence of the galaxy density on large-scale perturbations via all local gravitational observables.
The stochastic contributions, i.e.~the scatter around \refeq{dgdetL_intro}, remain much less studied. 
In \cite{Desjacques:2016bnm}, motivated by expressions in \cite{MSZ}, the stochastic part of the bias expansion was written as
\be
\d_g(\vx,\tau) = \dgdet + \epsold(\vx,\tau) +  \sum_O \epsold_O(\vx,\tau) O(\vx,\tau) \,.
\label{eq:dgstochL_oldintro}
\ee
with a set of noise fields $\{ \epsold, \epsold_O\}$.
Ref.~\cite{DAmico:2022ukl} employs a similar expansion in terms of multiple stochastic fields.
Each of those fields is completely characterized  by their local cumulants $\< (\epsold_O)^m(\vx)\>$. The fields $\epsold$ and $\epsold_O$ are considered first-order in perturbations.
We refer to this ansatz here as {\it non-minimal} noise theory.

Recently, Ref.~\cite{Rubira:2024tea} presented a general expression for galaxy statistics, at all orders in perturbation theory, at the partition function level. This partition function directly yields expressions for $n$-point correlation functions for any $n$ and at any loop order, including all stochastic contributions generated in the EFT. However, the result is quite abstract, and its connection to field-level expressions such as \refeq{dgstochL_oldintro} remained unclear. 
In this work, we show that the expression for the galaxy density \emph{field} can in fact be reduced to
\be
\d_g(\vx,\tau) = \sum_{m=0}^\infty \sum_{\unit, O} \beps{m}_O(\tau) \left[\epsG(\vx)\right]^m O(\vx,\tau) \,,
\label{eq:dgstochL_newintro}
\ee
with \emph{a single unit  Gaussian noise field} $\epsG(\vx) \sim \mathcal{N}(0,1)$, and a set of free coefficients $\{\beps{m}_\unit, \beps{m}_O\}$. 
Notice that the contributions with $m=0$ correspond exactly to the deterministic part, $\beps{0}_O = b_O$ in \refeq{dgdetL_intro}.\footnote{The contribution $\beps{0}_\unit$ multiplying the constant operator $\unit$ is usually dropped since it corresponds to a spatially constant contribution ($\vk=0$ mode).}
Conversely, the contributions with $m>1$ and $O = \unit$ generate the non-Gaussian stochasticity, i.e.~the higher cumulants of the noise. The contributions with $m>0$ and $O \neq \unit$ generate a modulation of the noise by large-scale perturbations. 
Here, we consider a single tracer $\d_g$; in case of multiple tracers, one would need to add an individual field $\epsG$ for each tracer (see e.g.~\cite{Hamaus:2010im,Rubira:2025scu,Voivodic:2025quw}). Further, we will drop the time argument from the fields and coefficients in most of the paper for clarity. Note that the field $\epsG(\vx)$ is explicitly time-independent, with all time dependence of stochasticity absorbed in the coefficients $\beps{m}_{O}$. 

This result represents a substantial simplification: while \refeq{dgstochL_oldintro} adds a \emph{non-Gaussian stochastic field for every bias operator}, \refeq{dgstochL_newintro} states that the stochastic part of the galaxy density field can be written in terms of a \emph{single Gaussian} stochastic field, and a set of generalized bias coefficients $\beps{m}_O(\tau)$.
The reason for this simplification is that statistics derived from \refeq{dgstochL_oldintro} still contain a number of degenerate contributions. This is shown at the level of galaxy $n$-point functions in \refapp{npt}, which also argues that \refeq{dgstochL_newintro} is sufficient to describe the non-degenerate stochastic contributions.

Our main focus in this paper however is the field-level likelihood for biased tracers.
We show that \refeq{dgstochL_newintro} enables a practical implementation of a field-level likelihood that can be used to incorporate all stochastic contributions in the EFT for field-level inference applications 
\cite{2013MNRAS.432..894J, 2013MNRAS.429L..84K, 2013ApJ...772...63W, Wang:2014hia, Modi:2018cfi, Schmidt:2020viy, Schmidt:2020tao, Shallue:2022mhf, Chen:2023uup, Andrews:2022nvv, Modi:2022pzm, Dai:2022dso, Kostic:2022vok, Qin:2023dew, Jindal:2023qew, Charnock:2019rbk, Doeser:2023yzv, Babic:2024wph, FBISBI}.
We emphasize that our goal is to obtain a likelihood that can capture all contributions, order by order, in the EFT of LSS. Other non-Gaussian likelihoods, such as Poisson or log-normal that have been used in empirical models for field-level inference, are not suited for this.
This generalized likelihood in the ``noise-field formulation'' proceeds by jointly inferring the initial density field $\din$ and the noise field $\epsG$, with independent Gaussian priors. Note however that the field $\epsG$ is an \emph{effective} (or ``nuisance'') field, not a physical field; in other words, the specific realization of $\epsG$ has no physical significance. This is in keeping with the fact that the initial conditions $\din$ are characterized by a single adiabatic growing mode, which we constrain with a single tracer density field $\d_g$. Multiple tracers $\d_g^a$ co-located in volume would require the introduction of multiple fields $\epsG^a$, as mentioned above.

We then present first results from an actual implementation of the noise-field formulation in the \leftfield\ code, in the form of field-level inference results on dark matter halos, considering the same sample as studied in \cite{FBISBI}.  That is, we extend the forward model adopted in \cite{FBISBI} to include non-Gaussian stochasticity.

\subsection{Outline of the paper} \label{sec:introsummary}  

We outline the main calculations of the paper here, providing a guide for the reader.

In \refsec{L}, we adopt the partition-function formulation for the noise from \cite{Rubira:2024tea}, in which higher powers of the current $J$ encode information about the non-Gaussian noise. We show how integrating out $J$ yields the likelihood. We then examine two limiting cases of this likelihood in \refsec{limiting}, and discuss the most general case, obtained by expanding terms of order $J^3$ and higher, in \refsec{currentmarginal_gaussian}. We discuss issues with this formal likelihood expansion, including negative probability densities, that preclude it from being used for field-level inference in practice.

We then turn to the ansatz \refeq{dgstochL_newintro}, which introduces a single Gaussian noise field with nonlinear couplings and generalized bias coefficients in \refsec{NS}. We demonstrate that this formulation leads to a field-level likelihood that is equivalent, order by order, to that obtained in \refsec{currentmarginal_gaussian} from the partition-function formulation. At the same time, this formulation avoids the issues of the formal likelihood expansion by essentially resumming higher-order terms to a positive-definite probability distribution.

In \refsec{results}, we present the first numerical implementation of field-level non-Gaussian noise, by introducing the noise as an additional noise field to be sampled. We show that this noise formulation produces more stable results than the previously considered Gaussian noise model. We conclude in \refsec{concl}.

\subsection{Notation}
\label{sec:notation}

We introduce the notation of the paper in this section. For the momenta integrals, we use
\be
\int_{\vp_1,\dots,\vp_n} = \int \frac{d^3p_1}{(2\pi)^3} \dots \int \frac{d^3p_n}{(2\pi)^3} \,.
\ee
We consider $W_\L$ sharp-in-$k$ filters at a scale $\L$ (and similarly $\kmax$) and write the filtered fields as
    \be \label{eq:filtering}
        f_{\L}(\vk) = \,W_\L(\vk) f(\vk)\,.
        \ee
The Gaussian linear density field $\dinL$ power spectra is given by
\be
\Plin^\L(k) = \< \dinL(\vk)\dinL(\vk')\>^{'},
\ee
where we use the prime to encapsulate the momenta conservation in the $n$-point functions
      \be
      \<O(\vk_1) \dots O(\vk_n)\> = \, \diracpi(\vk_{1\dots n})\,\<O(\vk_1) \dots O(\vk_n)\>'\,,
      \ee
with $\diracpi = (2\pi)^3\d_{\rm D}$ the (3D) Dirac delta distribution
The variance of the linear density field on a scale $\L$ is 
\be
\sigma^2_\L  = \int_{\vp}^\L \Plin(p)\,.
\label{eq:sigmaLambda}
\ee  
Throughout this work we assume initial conditions smoothed at a scale $\Lambda \gg k$, for all $k$ values for which observables are evaluated. To avoid cluttering the text, we remove the $\L$ subscript from the fields in the following sections, but it should be assumed the initial conditions are always smoothed on the scale $\L$.

 The bias operators are constructed on top of the linear fields via the convolution with a $K_O$ kernel
    \be
    O(\vk) = \sum_{n=n(O)}^\infty\int_{\vp_1,\ldots,\vp_n} \diracpi(\vk-\vp_{1\ldots n}) K_O^{(n)}(\vp_1,\ldots \vp_n) \din(\vp_1) \cdots \din(\vp_n) \,.
    \label{eq:Ok}
    \ee
where $n(O)$ is the leading perturbative order at which the operator $O$ contributes.

We always subtract the mean of nontrivial operators,
\be \label{eq:Onorm}
O \to O - \langle O\rangle \,.
\ee    
We consider the general bias expansion consisting of a set of operators $O$, ordered in perturbation theory and spatial derivatives. Our results are valid at all orders in perturbations, so we do not consider a fixed maximum order here.
Throughout, higher-derivative operators are also included, noting that they are controlled by $k^2/k_{\rm nl}^2$, where $k_{\rm nl}$ is the non-linear scale where perturbation theory fails, or $k^2 R_*^2$, where $R_*$ is the spatial length scale associated with the formation of the galaxies considered. We also consider the (zeroth-order) unit operator
\be \label{eq:unit}
\unit(\vk) = \diracpi(\vk)\,,
\ee
which only contributes to stochastic terms (higher-order-in-current operators) since the unit operator is removed from the bias basis by \refeq{Onorm} (equivalent to demanding $\langle \d_g \rangle = 0$).

In either the non-minimal or Gaussian-noise formulation presented above, the noise field $\eps$ is characterized by a two-point function given by
    \be \label{eq:hd_exp_noise}
    \langle \eps(\vx) \eps(\v{y}) \rangle = [P_{\eps,\unit} + P_{\eps,\nabla^2\unit} \nabla_{\vx}^2 + \ldots] \delta_{\rm D}(\vx-\v{y}),
    \ee
    corresponding to a local stochastic process. Here, we have also written the leading higher-derivative term to illustrate their structure. 
The field $\eps$ is by definition uncorrelated with the initial density field: 
\be
\< \eps(\vk) \din(\vk')\> = 0,
\quad\mbox{and moreover}\quad
\< \eps(\vk) \din(\vk_1) \cdots \din(\vk_n)\> = 0,
\ee
from which follows $\<\eps(\vk) O[\din](\vk')\> = 0$.
See \refapp{npt} for examples on how the noise field appears in $n$-point functions for both formulations \refeqs{dgstochL_oldintro}{dgstochL_newintro}.

\paragraph{Index notation.}

We widely use the index notation for fields in Fourier space:
\be
X(\vk_i) \to X^i\,,
\qquad\mbox{and} \qquad
X(-\vk_i) \to X^{-i}\,.
\ee
The integral over a field is written as 
\be
\int \Del X \equiv \prod_i \int dX^i\,.
\ee
We define the corresponding Dirac delta distribution by
\ba \label{eq:diracshort}
\diracpi(\vk_{j_1 \ldots j_m} + \vk_{i_1\ldots i_n}) \equiv \diracindex{i_1\ldots i_n}^{j_1 \ldots j_m} \,,
\ea
which will be useful notationally. Contracted indices indicate an integral, e.g.
\ba
  \diracindex{ij} X^i Y^j &=  X^i Y^{-i}  \equiv \int_{\vk} X(\vk) Y(-\vk)\,,  \vs
 \diracindex{ijk} X^i Y^j Y^k &\equiv \int_{\vk_1,\vk_2} X(\vk_1) Y(\vk_2) Y(-\vk_{12})\,.
  \ea
  With this, \refeq{dgdetL_intro} reads
\be
\dgdet^i[\{b_O\}, \din] = \sum_O b_O O^i[\din].
\label{eq:dgdetL}
\ee

\paragraph{Gaussian action.}    
The Gaussian free action or prior is given by the matter $2$-point function
    \be
        \P[\din] = \left(\prod_{\vk}^\L 2\pi \Plin(k)\right)^{-1/2} \exp\left[-\frac12 \int_{\vk}^\L \frac{|\din|^2}{\Plin(k)}\right]\,. \label{eq:PlikeDef}
    \ee
We drop any
prefactors in the likelihood that are independent of the parameters of
interest (in particular powers of $2\pi$ and such), since those are
irrelevant for the inference of parameters in the field $\din$. 
Using the index notation, we can write \refeq{PlikeDef} as 
    \be
    \P[\din] = \left(\prod_{\vk}^\L 2\pi \Plin(k)\right)^{-1/2} \exp\left[
      -\frac12 \din^i \left(\Plin^{-1}\right)_{ij} \din^j\right], \quad\mbox{with}\quad
    \left(\Plin^{-1}\right)_{ij} = [\Plin(k_i)]^{-1} \diracpi^{ij} \,. \label{eq:PlikeDefindex}
    \ee
Throughout, we will suppress time arguments, and often the dependence on cosmological parameters as well.

We will make much use of the well-known Gaussian integral identity
  \be
\int d^d X \exp \left[ -\frac{1}{2} \v{X}^\top \v{A} \v{X} +
      \v{B}^\top \v{X} \right] = \frac{(2\pi)^{d/2}}{|\v{A}|^{1/2}}
\exp \left[ \frac{1}{2} \v{B}^\top \v{A}^{-1} \v{B} \right] \, ,
\label{eq:gaussint}
\ee
and use bold-face to denote field-space objects with one or two indices,
in those cases where index contractions are obvious.

\paragraph{Kernel definition.}

We now introduce general kernels used in the paper, for reference. Since they make reference to concepts introduced later, they can be skipped at first reading.
We define
\ba \label{eq:Knotation}
  \K^{\{m\},j_1\ldots j_m}[\{\CO{m}_O\},\din] &\equiv  \sum_{\unit, O}  \CO{m}_O \sum_{n=n(O)}^\infty \diracindex{{i_1\ldots i_n}}^{j_1 \ldots j_m}  \ker^{(n)}_{O}(\vk_{i_1}, \ldots, \vk_{i_n})\din^{i_1}\cdots \din^{i_n} \,.
  \ea
Consistent with \refeq{dgstochL_newintro}, we use braces to denote the $m$-th order in the current or noise field, and parentheses for the $n$-th order in perturbation theory. 
 These kernels depend on a set of coefficients $\{\CO{m}_O\}$ at fixed $m$ which we will discuss in the following section. 
  Note that we always include the unit operator in the basis.
  We also define $\mu$ and $\Sigma$ for the cases $m=1,2$ respectively:
\ba\label{eq:KKdefm1}
\mu^j[\{b_O\},\din] &\equiv \K^{\{1\},j}[\{b_O\},\din]
 =  \sum_{\unit, O} b_O \sum^{\infty}_{n=n(O)} \diracindex{i_1\ldots i_n}^{j} \ker^{(n)}_{O}(\vk_{i_1}, \ldots, \vk_{i_n})\din^{i_1}\cdots \din^{i_n} \\
&=  \dgdet^{-j}[\{b_O\}, \din] = \sum_{O} b_OO^{-j}[\din] \,,\vs
 \Sigma^{jk}[\{\CO{2}_O\},\din] &\equiv \K^{\{2\},jk}[\{\CO{2}_O\},\din]  =  \sum_{\unit, O} \CO{2}_O \sum^{\infty}_{n=n(O)} \diracindex{i_1\ldots i_n}^{jk} \ker^{(n)}_{O}(\vk_{i_1}, \ldots, \vk_{i_n})\din^{i_1}\cdots \din^{i_n} \vs
&= \CO{2}_\unit \diracpi^{jk} + \diracindex{l}^{jk} \sum_O \CO{2}_O O^{l}[\din]\,.
\nonumber
\ea

\section{From the partition function to the likelihood}
\label{sec:L}

In this section, the goal is to obtain the EFT likelihood starting from the EFT partition function.\footnote{Strictly speaking, we will be deriving the posterior for cosmological and bias parameters given the galaxy density field, while the likelihood is part of the integrand, see \refsec{posterior_from_Z}. We continue to use the term ``likelihood'' here loosely, following the literature (e.g.~\cite{Cabass:2019lqx}).} We present the partition function in \refsec{stoch_theory} and calculate the likelihood in \refsec{posterior_from_Z}. 
In \refsec{limiting} we discuss some simplified cases, and,
in \refsec{currentmarginal_gaussian}, perform a formal expansion of the likelihood in terms of, essentially, moments of the galaxy stochasticity.

\subsection{The EFT partition function} \label{sec:stoch_theory}

In this section we review the EFTofLSS partition function including noise terms. In \cite{Rubira:2024tea} (see also \cite{Cabass:2019lqx, Cabass:2020nwf}), it was shown that the partition function can be written as 
\ba
Z[\v{J}] &= \int \Del\din \P[\din] \exp \left( S_{\rm eff}[\din, \v{J}] \right)\,,
\vs
\mbox{with}\qquad
S_{\rm eff}[\din, \v{J}] &\equiv \sum^{\infty}_{m=1} \sum_{\unit, O} \frac{\CO{m}_O}{m!} \int_{\vx} \left[J(\vx)\right]^m O[\din](\vx)\,.
\label{eq:ZofJ} 
\ea
Provided that the set $\{O\}$ is a complete set of linearly independent bias operators (see \cite{Desjacques:2016bnm}), this partition function can be shown to be closed under renormalization \cite{Rubira:2024tea} (see also \cite{Rubira:2023vzw, Nikolis:2024kbx,Bakx:2025cvu}). In the partition function, the terms with $m=1$ correspond to the usual bias parameters, $b_O = \CO{1}_{O}$, while the terms with $m\geq2$ correspond to stochastic contributions.
Notice that, for $m=1$, the zeroth-order $O = \unit$ is removed by \refeq{Onorm}.
Higher-derivative contributions for both the deterministic and stochastic fields are included in the partition function via current-derivative terms such as $\nabla^2J$ and $\partial_i J \partial^iJ$. We focus in this work on leading-in-derivative operators, but the results are straightforwardly generalized to include higher-derivative terms, including then derivative expansions of the noise field such as those in \refeq{hd_exp_noise}.

The relatively abstract expression \refeq{ZofJ} becomes more concrete when calculating $n$-point correlation functions, obtained by taking $n$ derivatives of the partition function with respect to the current $\v{J}$. Since
\ba
\langle \d_g(\vk_1) \dots \d_g(\vk_n) \rangle &= \int \Del\din \P[\din] \, \d_g(\vk_1) \dots \d_g(\vk_n) \, e^{S_{\rm eff}} \,,
\ea
the $n$-point function is given by
\be
\langle \d_g(\vk_1) \dots \d_g(\vk_n) \rangle =   \frac{1}{Z[\v{J}=\v{0}]}\frac{\delta^n Z }{\delta J(\vk_1) \dots \delta J(\vk_n)}\Big|_{\v{J}=0} \,.
\label{eq:Zderiv}
\ee
Therefore, as derived in \cite{Rubira:2024tea}\footnote{Ref. \cite{Rubira:2024tea} was lacking the Dirac factors with the $\CO{m}_\unit$ terms in their Eqs.~(2.8--2.10).}
\ba
\langle \d_g(\vk) \rangle &= \CO{1}_\unit \diracpi(\vk) = 0  \,,\vs
\langle \d_g(\vk_1) \d_g(\vk_2) \rangle &= \sum_O \sum_{O'} \CO{1}_O \CO{1}_{O'} \langle O(\vk_1) O'(\vk_2) \rangle + \CO{2}_\unit  \diracpi(\vk_{12})\,, \label{eq:dd_general}\\
\langle \d_g(\vk_1) \d_g(\vk_2) \d_g(\vk_3) \rangle &= \sum_O \sum_{O'} \sum_{O''} \CO{1}_O \CO{1}_{O'}\CO{1}_{O''} \langle O(\vk_1) O'(\vk_2)O''(\vk_3) \rangle \vs 
&\quad+ \left( \sum_O \sum_{O'} \CO{1}_O \CO{2}_{O'} \langle O(\vk_1) O'(\vk_{23}) \rangle + 2\,\textrm{perm.} \right)+  \CO{3}_\unit \diracpi(\vk_{123})\,.\nonumber
\ea
By simple dimensional analysis we have
\ba
[O(\vx)] &= d_O \quad\Rightarrow\quad  [O(\vk)] = d_O-3\,, \\
[J(\vx)] &= 3 \quad\Rightarrow\quad  [J(\vk)] = 0\,, \\
[\CO{m}_O] &= - \left[\int_{\vx} \v{J}^m O \right] = 3-3m-d_O \label{eq:dimC}\,.
\ea
where $d_O  = n_{\rm deriv}(O)$ is the number of derivatives in the operator $O$. 
We emphasize that the coefficients $\CO{m}_O$, defined in \refeq{ZofJ} are not the same as $\beps{m}_O$ introduced in \refeq{dgstochL_newintro}, the latter having dimension $-d_O$ while, for $d_O=0$, $\CO{m}_O$ have dimensions of an $m$-point function in Fourier space. We will return to the precise relation between the $\CO{m}_O$ and the $\beps{m}_O$ in the following section.  

Using the index notation, \refeq{PlikeDefindex}, and the kernels $\K$ introduced in \refeq{Knotation}, the partition function can also be  written as
\ba
Z[\v{J}|\{\CO{m}_O\}] 
  &= \int \Del\din \P[\din] \exp\bigg[  J_{i} \mu^{i}[\{b_O\},\din]  + \frac12 J_{i} J_{j} \Sigma^{ij}[\{\CO{2}_O\},\din] \vs
  &\hspace{3.6cm}  + \frac16 J_{i} J_{j} J_{k} \K^{\{3\},ijk}[\{\CO{3}_O\},\din] +\ldots \bigg]
\,.
\label{eq:ZL}
\ea
For more examples of the momentum structure of these terms, see \cite{Rubira:2024tea}. Note once more that $\mu^i = \K^{\{1\},i} = \dgdet^{-i}$ and that $\Sigma^{ij} = \K^{\{2\},ij}$.

\subsection{EFT likelihood from the partition function} \label{sec:posterior_from_Z}

Rather than $n$-point functions, we are interested here in
the ``likelihood'' for the galaxy density field. First, this is the crucial
ingredient in field-level inference approaches and for EFT-based generative
models, which are required for simulation-based inference, for example \cite{Tucci:2023bag,FBISBI}. Second, the field-level likelihood allows us to formulate field-level
expressions such as \refeqs{dgstochL_oldintro}{dgstochL_newintro} with more rigor, as it is not immediately obvious
how such relations follow from the partition function.

The \emph{posterior} for a set of cosmological parameters $\theta$ and EFT coefficients $\CO{m}_O$ given some observed data $\hdg$ is given by
\ba \label{eq:Plikely}
\P\left[\{\theta\},\, \{ \CO{m}_O \} \ |\ \hdgkmax\right] &\propto\\
&\hspace{-2cm}\int \Del\din \P[\din|\{\theta\}]\ \LL_{\kmax}[\hdgkmax | \din, \{\theta\}, \{\CO{m}_O\}]
\  \P(\{\theta\},\ \{\CO{m}_O\})\,, \nonumber
\ea
where $\LL_{\kmax}$ is the \emph{likelihood} proper, which depends on $\din$ (including modes up to $\L$)
through the bias operators $O[\din]$, and the subscript $\kmax$ indicates that only
modes in the data up to $\kmax$ are included. Following \cite{Schmidt:2025iwa}, this
corresponds to cutting all external momenta of $n$-point functions of the data
at $\kmax$. 
Finally, $\P(\{\theta\},\ \{\CO{m}_O\})$ denotes the prior on
EFT and cosmological parameters.

Our main goal in this section is to connect the likelihood $\LL$ to the partition function. In order to keep the equations simple, we will drop the prior
$\P(\{\theta\},\ \{\CO{m}_O\})$ in the following, and also keep only the
dependence of the data in the posterior explicit, defining
\be
\hatP[\hdgkmax]
\equiv \int \Del\din \P[\din|\{\theta\}] \ \LL_{\kmax}[\hdgkmax | \din, \{\theta\}, \{\CO{m}_O\}]\,,
\label{eq:hatPdef}
\ee
using a prime to differentiate from \refeq{Plikely}. Once an explicit expression for $\hatP$ is obtained, we
can then read off the likelihood proper, $\LL_{\kmax}$, from the integrand. 

The quantity $\hatP[\hdgkmax]$ represents the probability of finding the observed tracer field $\hdgkmax$, given (implicit) fixed values for the cosmological and EFT parameters $\theta,\  \CO{m}_O$ and including modes up to $\kmax$, marginalized over the entire initial conditions $\din$ (up to $\L$). Following \cite{Cabass:2019lqx}, this probability can be written as a field-level Dirac delta functional, which we in turn represent in Fourier space:
\ba
\hatP[\hdgkmax] &=
\int \Del \dgkmax \P[\dgkmax] \dirac{0,\kmax}(\hdgkmax-\dgkmax)  \vs
&=  \int  \Del (\ii^{-1}\v{J}_{\kmax}) \left\< \exp\left[J_{\kmax}^{i} (\hdg-\dg)_{-i} \right]\right\> \vs
&=  \int \Del (\ii^{-1}\v{J}_{\kmax}) \exp\left[J_{\kmax}^{i} (\hdg)_{-i} \right]
\left\< \exp\left[-J_{\kmax}^i (\dg)_{-i} \right]\right\> \vs
&=   (Z[\v{J}=\v{0}])^{-1} \int \Del (\ii \v{J}_{\kmax}) \exp\left[-J_{\kmax}^{i} (\hdg)_{-i} \right] Z[\v{J}_{\kmax}]\,.
\ea
Here, $\hdg$ represents the point in field space at which we evaluate the probability $\hatP$ (e.g., the observed data), while $\dg$ represents the random field itself.
We denote the field-level Dirac delta that includes all modes in the field up to $\kmax$ (except for the zero mode) with $\dirac{0,\kmax}$.
In the second line, we have used the Fourier representation of the Dirac delta in field space, and expressed the integral over $\dgkmax$ as expectation value.
That is, the expectation values in the second and third lines are taken with respect to the measure $\P[\din|\{\theta\}]$, just as for $n$-point correlators.
Note that we have chosen an imaginary current $\v{J}$ here for convenience, so that the integral measure is accompanied by a factor $\ii^{-1}$,
and changed the sign of $\v{J}$ in the last line.
We have emphasized that the current $\v{J}$ employed here likewise only has support up to $\kmax$, and that $Z[\v{J}=\v{0}]$ is the partition function evaluated at zero current, which still has a nontrivial dependence on the parameters $\{\CO{m}_O\}$, implicit here but important to obtain a normalized probability distribution. 
The last equality can be shown via a formal Taylor expansion
\ba
\< \exp\left[J_{\kmax}^{i} (\dg)_{-i} \right]\> &= \sum_{l=0}^\infty \frac1{l!}
J_{\kmax}^{i_1}\cdots J_{\kmax}^{i_l} \< (\dg)_{-i_1} \cdots (\dg)_{-i_l} \>
\vs
&= (Z[\v{J}=\v{0}])^{-1} \sum_{l=0}^\infty \frac1{l!}
J_{\kmax}^{i_1}\cdots J_{\kmax}^{i_l} 
\frac{\Del}{\Del J_{\kmax}^{i_1}} \cdots \frac{\Del}{\Del J_{\kmax}^{i_l}} Z[\v{J}_{\kmax}] \Big|_{\v{J}_{\kmax} = \v{0}} \vs
&= (Z[\v{J}=\v{0}])^{-1} Z[\v{J}_{\kmax}].
\ea
Reinstating the explicit dependencies [but dropping the prior $\P(\{\theta\},\ \{\CO{m}_O\})$], this becomes
\ba
\P\left[\{\theta\},\, \{ \CO{m}_O \} \ |\ \hdgkmax\right] &\propto \label{eq:PgivenZ} \\
&\hspace{-2cm}
   \frac{1}{Z[\v{0}|\{\theta\},\ \{\CO{m}_O\}]}\int \Del (\ii \v{J}_{\kmax}) \exp\left[- J_{\kmax}^{-i} (\hdgkmax)_{i} \right] Z[\v{J}_{\kmax}|\{\theta\},\ \{\CO{m}_O\}]\,.
\nonumber
\ea
\refeq{PgivenZ} states that the EFT posterior is obtained as the functional Fourier transform of the partition function, in agreement with \cite{Cabass:2019lqx}, but making the cutoff $\kmax$ explicit.
Notice that, since the current is cut at $\kmax$, no modes above this value are excited in the partition function. Thus, the likelihood is related to the partition function in precisely the same regime as $n$-point functions up to the same $\kmax$.
Finally, note that we have kept the normalization $Z[\v{0}|\{\theta\}, \{\CO{m}_O\}]$, as it is parameter-dependent and hence contributes nontrivially to the posterior in $\{\theta\}, \{\CO{m}_O\}$ (any parameter-independent constants on the other hand can safely be ignored, as we are not attempting to compute the normalizing evidence).
In the following, we will neglect the dependence on the cosmological parameters $\{\theta\}$, since it no longer plays any role in the derivation.

\subsection{Limiting cases} \label{sec:limiting}

While we have formally obtained the likelihood as \refeq{PgivenZ}, the integration over $\v{J}$ in \refeq{PgivenZ} is not tractable for the general $Z[\v{J}]$ in \refeq{ZofJ}. 
In the following, we will discuss simpler limiting cases for which the integration can be done analytically. While in \refsec{currentmarginal_m2} we truncate the $\v{J}$ series keeping only terms up to $\v{J}^2$, in \refsec{currentmarginal_O1} we keep all powers in $\v{J}$ but neglect the coupling with $\din$. Finally, in \refsec{currentmarginal_gaussian} we consider an expansion in the current, which is equivalent to expanding around the case considered in \refsec{currentmarginal_m2}.

\subsubsection{Quadratic-in-current terms and all operators}
\label{sec:currentmarginal_m2}

We start by restricting the partition function \refeq{ZofJ} to terms up to second-order in the current $\v{J}_{\kmax}$ and including the most general set of bias operators $O[\din]$.
Inserting that into \refeq{PgivenZ} yields
\ba
\hatP[\hdgkmax] 
&=  (Z[\v{0}|\{b_O, \CO{2}_O\}])^{-1} \int \Del (\ii \v{J}_{\kmax}) \exp\left[- J_{\kmax}^{i} (\hdg)_{-i} \right]
   \label{eq:PtoX2} \\
  %
   \times &\int \Del\din \P[\din]\exp\bigg[J_{\kmax,i} \mu^{i}[\{b_O\},\din]    + \frac12  J_{\kmax,i} J_{\kmax,j} \Sigma^{ij}[\{\CO{2}_O\},\din] \bigg] \nonumber
\,,
\ea
where we have used $\mu$ defined in \refeq{KKdefm1}, and
\be
\Sigma^{ij}[\{\CO{2}_O\}, \din] = \CO{2}_\unit \diracpi^{ij} + \diracindex{k}^{ij} \sum_O \CO{2}_O O^{k}[\din]\,.
\label{eq:KKdef}
\ee
We omit the higher-derivative terms in \refeq{hd_exp_noise}, but one can straightforwardly generalize the conclusions of this section to include them.

In this case we can perform the Gaussian integral over $\v{J}_{\kmax}$ in \refeq{PtoX2} using \refeq{gaussint}, yielding
\ba 
\hatP[\hdgkmax] 
=\:& \int \Del\din \P[\din]
\LL_{\kmax}[\hdgkmax | \din, \{\theta\}, \{\CO{1,2}_O\}]
\nonumber
\ea
with
\ba
\LL_{\kmax}[\hdgkmax | \din, \{\theta\}, \{\CO{1,2}_O\}]
=\:& \N_{\LL}[\{\CO{2}_O\},\din] \label{eq:Pgaussgen}\\
 &
\hspace{-3cm} \times\exp\bigg[ Y^i[\hdgkmax,\{b_O\},\din]\left(\Sigma^{-1}[\din,\{\CO{2}_O\}]\right)_{ij} Y^j[\hdgkmax,\{b_O\},\din] \bigg] \,,
\nonumber
\ea
and 
\be \label{eq:Ydef} 
Y^i[\hdgkmax,\{b_O\},\din] \equiv \hdgkmax^{-i} - \d_{g, {\rm det}}^{-i}[\{b_O\},\din] =\hdgkmax^{-i} - \mu^{i}[\{b_O\},\din] \,,
\ee
which we will widely use below, as the likelihood becomes centered in terms of $Y$.
Note that the coupling with $\v{J}_{\kmax}$ in \refeq{PtoX2} ensures that both $\hdg$ and $\mu$ are cut at $\kmax$, hence we will always only encounter the filtered $\v{Y} = \v{Y}_{\kmax}$,  and throughout we omit $\kmax$. 
Further, we defined the likelihood normalization
\be
\N_{\LL}[\{\CO{2}_O\},\din] \propto (Z[\v{J}=\v{0}])^{-1} |\v{\Sigma}[\din]|^{-1/2} \,.
\ee
In the following, we will drop the explicit parameter dependence in $\mu$, $\Sigma$, and $Y$ for clarity.

It is clear that the likelihood (proper) in \refeq{Pgaussgen} can be interpreted as being due to a single, Gaussian stochastic degree of freedom $\eps^i$ with covariance $\Sigma^{ij}[\din]$.
The evaluation of this likelihood however
requires inverting the matrix $\Sigma$ and evaluating its determinant. 
If we keep only the first, diagonal term in \refeq{KKdef}, $\Sigma^{ij}$ is easily inverted; in fact,
\refeq{Pgaussgen} then becomes the Fourier-space likelihood first derived in \cite{Schmidt:2018bkr}, where the data $\hdgkmax$ and model prediction \refeq{dgdetL} are compared with a diagonal, Fourier-space covariance up to $\kmax$. 
Going beyond the leading term in \refeq{KKdef} however, i.e. when attempting to include the coupling between $\din$ and noise, this matrix can no longer be inverted analytically in general. Moreover, it is a dense matrix, which renders a numerical evaluation of the likelihood \refeq{Pgaussgen} essentially intractable.\footnote{In a field-level analysis, the likelihood has to be evaluated at every step in a sampling process. Since the matrix $\Sigma^{ij}$ has dimension $D\times D$, where $D$ is the total dimension of the density field $D = N_{\rm grid}^3$, a matrix inversion is impractical in reality. In addition, a matrix of this size cannot be stored in memory.}
An alternative route was proposed in \cite{Cabass:2020nwf}, by instead formulating the likelihood in real space. This approach is unfortunately hampered by the need to enforce the Fourier-space cut at $\kmax$. We discuss this in \refapp{reallikelihood}. 

These issues aside, \refeq{Pgaussgen} is not complete in any case, as we have truncated the partition function at order $\v{J}^2$. We turn to the higher-order terms in $\v{J}$ next.

\subsubsection{All-orders-in-current terms but only unit operator $O = \unit$} \label{sec:currentmarginal_O1}

Let us now study a second instructive case, the limit in which we keep only the coefficients $\CO{m}_\unit$, i.e.~the subset of kernels in \refeq{Knotation}
given by $\CO{m}_\unit \diracpi^{j_1\dots j_m}$ for $m\geq 2$.
That is, we keep arbitrary powers of the current, but neglect the coupling with $\din$ in the stochastic ($m \geq 2$) contributions, while keeping the full set of deterministic ($m=1$) terms. 
At the level of $n$-point functions, this corresponds to keeping only the purely stochastic ``shot-noise'' terms in \refeq{dd_general}.

In this limiting case, the posterior is given by
\ba
\hatP[\hdgkmax] 
&=  (Z[\v{J}=\v{0}])^{-1} \int \Del (\ii \v{J}_{\kmax}) \!\!
\int \Del\din \P[\din] \exp\bigg[
-  J_{\kmax,i} Y^i[\hdgkmax,\{b_O\},\din]  \vs
  &  \hspace{3cm}+ \sum_{m=2}^\infty \frac{1}{m!} \CO{m}_\unit \diracpi^{j_1\dots j_m}  J_{\kmax,j_1}\cdots J_{\kmax,j_m}  \bigg]
\,, \label{eq:P_limiting_step0}
\ea
which we can write as
\ba
\hatP[\hdgkmax] &= \int \Del\din \P[\din] 
\LL_{\kmax}\left[\hdgkmax - \mu_{\kmax}[\{b_O\},\din],\ \{\CO{m\geq 2}_\unit\} \right]\,,
\ea
where
\ba
\LL_{\kmax}\left[Y,\ \{\CO{m\geq 2}_\unit\}\right] &\equiv (Z[\v{J}=\v{0}])^{-1}  \int \Del (\ii \v{J}_{\kmax}) \label{eq:Pstoch1}\\
& \hspace{1.5cm}\times\exp\bigg[- J_{\kmax}^{i} Y_i  + \sum_{m=2}^\infty \frac{1}{m!}  \CO{m}_\unit \diracpi^{j_1\dots j_m}  J_{\kmax,j_1}\cdots J_{\kmax,j_m}  \bigg]\,. \nonumber
\ea
\refeq{Pstoch1} shows explicitly that, also for this limiting case, a single noise field with PDF $\LL_{\kmax}[\cdot]$ captures all stochastic contributions $\CO{m}_\unit$ to the tracer density field. 
Moreover, we can identify the coefficients $\CO{m}_\unit$ as the cumulants of the noise field (recall that $Y$ is the residual between the data $\hdgkmax$ and mean-field prediction $\mu_{\kmax}$). 
Therefore, the partition function, with the linear-in-current term factored out, corresponds to the \emph{characteristic functional} of the stochastic process.
Notice that the contraction between $Y$ and $J_{\kmax}$ implies that $\LL_{\kmax}[Y]$ does not depend on momenta in $Y$ that are beyond $\kmax$. 

Unfortunately, even if the coupling of stochasticity and initial conditions could be neglected as done here, \refeq{Pstoch1} is not very practical. The EFT expansion demands that we employ a non-Gaussian probability distribution that allows for a flexible specification of an arbitrary number of higher-order cumulants (skewness, kurtosis, ...), depending on the order of the expansion considered, but no such closed-form distribution exists. Nevertheless, \refeq{Pstoch1} could be interesting as a starting point for a resummation of noise contributions to an approximate, closed-form non-Gaussian PDF.

\subsection{General expansion around the Gaussian likelihood} \label{sec:currentmarginal_gaussian}

So far, we have considered the likelihood in the limits of only $\CO{\leq 2}_O \neq 0$ (\refsec{currentmarginal_m2}), and only $\{ \CO{1}_O,\  \CO{m}_\unit \}\neq 0$ (\refsec{currentmarginal_O1}), respectively. In both cases, we could arrive at all-order expressions for the likelihood, and we found that all noise terms could be captured by a single non-Gaussian stochastic field (with the cumulants given by $\CO{m}_\unit$ in the latter case).

In this section, we push the results from \refsec{currentmarginal_m2} further, by perturbatively including the higher-in-current terms without neglecting the coupling to $\din$.  
Concretely, we perform a perturbative expansion in the current $\v{J}$, which corresponds to an Edgeworth-like expansion in the cumulants of the stochastic field.
We emphasize that the expressions are valid at all orders in $\din$. 
We can write
\ba
\hatP[\hdgkmax] 
&=  (Z[\v{J}=\v{0}])^{-1} \int \Del (\ii \v{J}_{\kmax}) 
\int \Del\din \P[\din] \exp\bigg[  -J_{\kmax,i} Y^i  + \frac12 J_{\kmax,i} J_{\kmax,j} \Sigma^{ij}\bigg] \vs
&\hspace{2cm} \times\exp\bigg[ \sum_{m=3}^\infty \frac{1}{m!} \K^{\{m\},j_1\ldots j_m} J_{\kmax,j_1} \cdots J_{\kmax,j_m} \bigg]
\vs
&=  (Z[\v{J}=\v{0}])^{-1} \int \Del\din \P[\din] \int \Del (\ii \v{J}_{\kmax}) 
 \exp\bigg[  -J_{\kmax,i} Y^i  + \frac12 J_{\kmax,i} J_{\kmax,j} \Sigma^{ij}\bigg] \vs
&\hspace{2cm} \times\bigg[ 1 + \sum_{L=3}^\infty \frac{1}{L!} \C^{\{L\},j_1\ldots j_L} J_{\kmax,j_1} \cdots J_{\kmax,j_L} \bigg]
\,,  \label{eq:Pexpand}
\ea
where in the second equation we expanded the exponential and we have defined
 (see \refapp{derivations} for the complete derivation)
  \ba
\C^{\{L\},j_1\ldots j_L}[\{\CO{\geq 3}_O\},\din]
&= \sum_{\substack{a_3,\cdots,a_L \geq 0\\ 3 a_3 + \cdots + L a_L = L} }
\frac{L!}{a_3! \cdots a_L!}
\prod_{m=3}^L\frac{1}{(m!)^{a_m}}\left(\K^{\{m\} \cdots}[\{\CO{m}_O\},\din]\right)^{a_m}
\label{eq:CCdef}
\,,
\ea
such that $\C^{\{L\},j_1\ldots j_L}$, appearing at order $\v{J}^L$, contains products of the kernels $\K^{\{m\}}$ (and the contributions are correspondingly multiplied by products of $\CO{m}_O$). For examples, see \refeq{C_examples}.

Our aim is to perform the $\v{J}$ integral at fixed $\din$.
In the following, we drop the arguments and the $\kmax$ subscript for clarity. 
We now shift the integration variable as\footnote{Note that,in Fourier space, $J$ has dimension 0, $Y$ has dimension $-3$, $\Sigma^{-1}$ has dimension 6, and the integration over $j$ has dimension $3$.}
\ba
J^i \to \tilde J^i &= J^i -\left(\v{\Sigma}^{-1}\right)^i{}_j Y^j.
\ea
This completes the square in \refeq{Pexpand}, and yields 
\ba
\hatP[\hdgkmax] 
 =  (Z[\v{J}=\v{0}])^{-1} \int& \Del\din \P[\din]  \int \Del (\ii \v{\tilde J}) 
\exp\bigg[-\frac12\v{Y}^T \v{\Sigma}^{-1} \v{Y} + \frac12 \v{\tilde{J}}^T \v{\Sigma} \v{\tilde{J}} \bigg] \vs
& \hspace{-2cm}\times\bigg[ 1 + \sum_{L=3}^\infty \frac{1}{L!} \C^{\{L\},j_1\ldots j_L}
  (\v{\tilde{J}} + \v{\Sigma}^{-1} \v{Y} )_{j_1} \cdots (\v{\tilde{J}} + \v{\Sigma}^{-1} \v{Y} )_{j_L} \bigg]
\,.
\label{eq:PexpandInt}
\ea
Now we have a sum over Gaussian integrals over $\tilde J$, which can be done to
yield (see \refapp{derivations})
\ba
\hatP[\hdgkmax] = \N_{\LL}
 \int \Del\din \P[\din]& \exp\left[ -\frac12  \v{Y}^T \v{\Sigma}^{-1} \v{Y} \right]
\bigg[ 1 + \sum_{m=3}^\infty \frac1{m!} \tilde{\C}^{\{m\},j_1\ldots j_m}
   Y_{j_1} \cdots Y_{j_m} \bigg]
\,,
\label{eq:PexpandFormal}
\ea
where the coefficients
\ba \label{eq:Ctilde_main}
\tilde\C^{\{m\},j_1\ldots j_m} =  \sum_{\substack{L=m\\ L-m\ \rm even}}^\infty 
\frac{L!}{2^{(L-m)/2} (L-m)!}
 \Csym^{\{L\},l_1\ldots l_L}& \left(\v{\Sigma}^{-1}\right)_{l_{m+1} l_{m+2}} \cdots \left(\v{\Sigma}^{-1}\right)_{l_{L-1} l_{L}} \vs
&\hspace{-2cm}\times (\v{\Sigma}^{-1})_{l_1 j_1} \cdots (\v{\Sigma}^{-1})_{l_m j_m}
\ea
involve contractions of the fully-symmetrized versions $\Csym^{\{L\geq m\}}$ [\refeq{Csym}] with powers of $\v{\Sigma}^{-1}$.
Note that
\be
\v{Y} = \v{Y}[\hdgkmax, \{b_O\},\din];\qquad \v{\Sigma} = \v{\Sigma}[\{\CO{2}_O\},\din]; \qquad
\tilde\C^{\{m\}} = \tilde\C^{\{m\}}[\{\CO{\geq 2}_O\}, \din]\,.
\ee
As in \refsec{currentmarginal_O1}, the expansion in $J$ of the partition function leads to an expansion in $Y$ of the likelihood. This expansion is valid if higher cumulants of the residual $Y$ between data and mean-field prediction are suppressed. This is the case if the noise is perturbatively close to Gaussian (see also the discussion in \cite{Cabass:2019lqx}).
The leading contribution to \refeq{Ctilde_main}, in the sense of the expansion discussed, is given by $L = m$, such that every $\CO{m}_O$ has a unique, leading-order contribution to $\tilde{C}^{\{m\}}$ with distinct shape, indicating that the $\CO{m}_O$ yield non-degenerate contributions to the field-level likelihood, just as they do in the partition function \refeq{ZofJ}.
Note that this equation also reproduces \refeq{Pgaussgen} if $\tilde{\C}^{\{m\}} = 0$ ($m\geq 3$), which holds of course if $\C^{\{L\}} = 0$ for $L\geq 3$.

\refeq{PexpandFormal} yields a consistent perturbative expansion in the EFT context, however it is not practical for real inference applications. The reason is twofold. First, the inversion of $\Sigma$ still presents the same obstacle as discussed in \refsec{currentmarginal_O1}. Second, when truncating at finite order $m$, the probability density becomes ill-defined (negative) in some regions of parameter space, and is not normalizable in general. This is typical of Edgeworth-like expansions in terms of cumulants.\footnote{The Edgeworth expansion in terms of cumulants $\kappa_\ell$ can be written as
\be \P(\d) = \frac{e^{-\d^2/2\sigma^2}}{\sqrt{2\pi}\sigma} \left[1 + \sum_\ell \kappa_\ell \frac{1}{\ell!\sigma^\ell}H_{\ell}(\delta) \right]\,,
\ee
with $H_\ell$ being the Hermite polynomials. In general, when truncated at any finite order $\ell$, there are regions in $\d$-space that have negative probability density.}
While higher-order terms in $m$ should in principle be suppressed following the discussion above, a numerical sampler will explore the tails of the distribution, and eventually encounter the regime of ill-defined probability. 
To conclude, the EFT likelihood \refeq{PgivenZ} is difficult to evaluate for the most general noise contributions.

However, we have seen in \refsec{limiting} that the stochastic contributions can be captured by {\it a single degree of freedom} both when restricted to the Gaussian case (but with mean and variance both coupled to $\din$) and the non-Gaussian case (with moments other than the mean not coupled to $\din$).
The generalization when including subleading noise contributions, as we just saw, indicates that one can consider expansions around a Gaussian field. 
In the next section, we will build on this to develop an alternative route to evaluating the EFT likelihood in the fully general case, with the goal
of obtaining a formulation that does not suffer from the above-mentioned problems, but yields an expansion that is perturbatively equivalent to \refeq{PexpandFormal}.

\section{The noise-field formulation of stochasticity}
\label{sec:NS}

As we have seen in the previous section, the formal likelihood obtained directly from the partition function is difficult to evaluate numerically and may lead to an ill-defined probability density when expanded in the cumulants (corresponding to an expansion in powers of the current in the partition function). In this section we take a different approach, starting instead with a field-level ansatz for the noise. Our goal is to show that this field-level ansatz is equivalent to the partition function approach, order by order in perturbation theory.
First, this provides a crucial conceptual clarification of galaxy stochasticity,
namely the statement that it is described precisely and unambiguously by
\refeq{dgstochL_newintro}, rather than \refeq{dgstochL_oldintro}. 
Second, thanks to the straightforward field-level formulation, this approach can be
directly used in field-level inference analyses to incorporate coupled and
non-Gaussian stochasticity at any order. We present results
in \refsec{results}.

\subsection{Definition}

Let us repeat the model of \refeq{dgstochL_newintro}. In the following, we will shorten $\epsG\to\eps$ for clarity of notation, since all instances of $\eps$ in this section correspond to the single Gaussian noise field introduced in \refeq{dgstochL_newintro}:
\ba
\dg(\vx,\tau) &= \dgdet + \sum_{m=1}^\infty \sum_{\unit,O}  \beps{m}_O(\tau) \left[\eps(\vx)\right]^m  O(\vx,\tau)\,, \vs
\mbox{with}\quad \dgdet &= \sum_{O}  \beps{0}_O(\tau)  O(\vx,\tau)  \,.
\label{eq:dgstochL_newrepeat}
\ea
We again keep the smoothing of the initial fields implicit, but in practice one would typically choose the same filtering scale $\L$ for both $\eps$ and $\din$.
Precisely, the joint prior on $\eps$ and $\din$ is given by
\be
\P[\din,\  \eps] = \mathcal{N}[\bm{0}, {\rm diag}\{\Plin(k)\}][\din]\times
\mathcal{N}[\bm{0}, {\rm diag}\{P_\eps\}][\eps]\,,
\label{eq:priordineps}
\ee
where $P_\eps$ is the constant power spectrum of the Gaussian field $\eps$. The numerical value of $P_\eps$ depends on the normalization convention which we leave unspecified here (see \refsec{results} for the lattice implementation).
The noise field $\eps$ can be understood as a ``nuisance field'' that captures how small-scale stochastic fluctuations affect the large-scale galaxy density field.
In the following, we will again drop the time arguments as we did after \refeq{dgstochL_newintro}.

The coefficients $\beps{m}_O$ in \refeq{dgstochL_newrepeat} are not the same as the $\CO{m}_O$.
While the dimensionless $\beps{m}_O$ are defined via the field-level formulation \refeq{dgstochL_newrepeat} and the index $m$ corresponds to the number of contracted stochastic fields (starting at zero, for the deterministic part), the dimensionful $\CO{m}_O$ are defined by the partition function of \refsec{stoch_theory} and $m$ corresponds to the number of contracted currents $J$ (starting at $m=1$ for the deterministic part). 
Specifically, we have 
\be \label{eq:beps0toCO1}
\beps{0}_O  = \CO{1}_O
\ee
for the usual deterministic bias coefficients, while the Gaussian stochasticity contribution to the galaxy power spectrum is at leading order described by 
\be\label{eq:b1toCO2}
(\beps{1}_\unit)^2 P_\eps = \CO{2}_\unit\,,
\ee
respectively in the two formulations. The leading non-Gaussian stochasticity is controlled by $\beps{2}_\unit$ vs. $\CO{3}_\unit$. Generally, the term $\propto \CO{m}_O$ in the partition function is indeed captured by $\beps{m-1}_O$ in the noise-field formulation, but with additional corrections.

We can now rewrite \refeq{dgstochL_newrepeat} using the index notation \refeq{Knotation} and \refeq{KKdefm1} as
\ba
\dg^{-k} &= \dgdet^{-k} + \sum_{m=1}^\infty \sum_{\unit, O} \beps{m}_O \sum_{n=n(O)}^\infty \diracindex{{i_1\ldots i_n}}^{k j_1 \ldots j_m}  \ker^{(n)}_{O}(\vk_{i_1}, \ldots, \vk_{i_n})\din^{i_1}\cdots \din^{i_n} \eps_{j_1}\cdots  \eps_{j_m}  \vs
&= \dgdet^{-k} +  \sum_{m=1}^\infty \K^{\{m+1\},kj_1\ldots j_m}[\{\beps{m}_O\},\din] \, \eps_{j_1}\cdots  \eps_{j_m} \,,
\label{eq:nsmodel}
\ea
where the deterministic part is given by $m=0$. Note that $\eps(\vk)$ has the same dimensions as as $\dgdet(\vk)$ and $\din(\vk)$.
The kernels here are precisely the same as those in the expansion of the partition function in $\v{J}$, we essentially just change coefficients from $\{\CO{m}_O\}$ to $\{\beps{m-1}_O\}$.\footnote{Note that $\K^{\{m\}}[\{\beps{m}_O\}]$ and $\K^{\{m\}}[\{\CO{m}_O\}]$ inherit different dimensionality from the coefficients $\beps{m}_O$ vs $\CO{m}_O$.}
Correspondingly, the contribution $\sim \beps{m}_O$ involves a kernel with $m+1$ upper indices, analogous to the contribution to $Z[\v{J}]$ controlled by $\CO{m+1}_O$.
We point out that the recent Ref.~\cite{Voivodic:2025quw} considered a similar expansion; we discuss the relation further below.

\subsection{Expansion around the Gaussian likelihood}

After having written down the field-level formulation \refeq{nsmodel}, we can use the Gaussian prior (or free action) \refeq{priordineps} for $\din$ and $\eps$ to express the PDF of $\dg$ in general as 
\ba
\hatP[\hdgkmax] 
= \int \Del\din \P[\din] &\N_\eps[P_\eps]\int \Del\eps 
 \exp\left[ -\frac12  P_\eps^{-1} \v{\eps}^T \v{\eps} \right] \vs
& \hspace{-2cm}\times \dirac{0,\kmax}\left(Y^k[\hdgkmax, \{\beps{0}_O\},\din] - \sum_{m=1}^\infty \K^{\{m+1\}, kj_1\ldots j_m}[\{\beps{m}_O\},\din] \eps_{j_1}\cdots  \eps_{j_m} \right)
\,,
\label{eq:nslike1}
\ea
where the variable $Y$ is defined in \refeq{Ydef}, and $\N_\eps[P_\eps]$ is the normalization of the prior over $\eps$ (fixed, since $P_\eps$ is kept fixed). We have used $k$ as placeholder index inside the field-level Dirac delta, in order to make the index structure clear.
This expansion in $\eps$ is valid as long as the dimensionless reduced cumulants $\kappa_n^{\eps} \equiv \< \eps^n \>_c / \< \eps^2 \>^{n/2}$ of the total noise contribution in \refeq{nsmodel} are much less than unity. Focusing on $\beps{m}_\unit$, and evaluating the corresponding leading-order contribution to the $(m+1)$-th reduced cumulant $\kappa_{m+1}^\eps$ (e.g. skewness for $m=2$, kurtosis for $m=3$, and so on), we find that
  \be
\kappa_{m+1}^\eps \stackrel{\rm LO}{=} \beps{m}_\unit (m+1)!\: \< \eps^2 \>^{(m-1)/2} \,.
  \ee
  Thus, assuming that $\beps{m}_\unit \sim 1$, we have an additional independent expansion parameter, which is the variance of the Gaussian noise field $\eps$, in analogy with how the variance of $\d$ controls the standard perturbative expansion. Both can be made sufficiently small by reducing $\kmax$.\footnote{Note that for interesting cosmological constraints, one typically considers tracers with significant signal-to-noise, i.e. for which $\< \eps^2 \> \lesssim \< \d^2 \>$.}

Our goal now is to integrate out the noise field $\eps$, and make a connection to the
expanded general PDF derived in the previous section, \refeq{PexpandFormal}.
The idea is that we can circumvent the complications found in \refsec{L} of terms nonlinear in the current by introducing the effective (Gaussian) noise field $\eps$.

We solve the integral over $\eps$ recursively in $m$, inserting the solution for the zero of the Dirac delta at each order into the next-order calculation.
We start by setting $m=1$, i.e. considering only the first, linear-in-$\epsilon$ term in the sum inside the
Dirac delta  in \refeq{nslike1}.
For this, we restrict the $\vk$ support of $\eps$ to $(0,\kmax)$, the
  same range as for the current $\v{J}_{\kmax}$ in \refsec{currentmarginal_gaussian}, and correspondingly restrict the matrix $\Sigma$ to this range.
Then, the Dirac delta fixes the solution for $\eps$ at linear order. 
  If we were to allow for higher $\vk$ support of $\eps$, some modes in $\eps$ would remain
  unconstrained; however, we expect that the effect of integrating out these
  additional modes would simply shift the coefficients $\beps{m}_{\unit,O}$.
  We obtain
\ba
\epslin^k[\hdgkmax, \{\beps{0}_O\},\{\beps{1}_O\},\din] &\equiv \left[\left(\K^{\{2\}}[\{\beps{1}_O\},\din]\right)^{-1}\right]_{kj} Y^{j}[\hdgkmax,\{\beps{0}_O\},\din]\,.
\ea
Note that the arguments of $Y$ are the same as in \refsec{L}, since they refer to the deterministic component described by the bias parameters proper ${\beps{0}_O}$ [see \refeq{beps0toCO1}]. In contrast, the arguments of $\K^{\{2\}}$ differ: while the partition function formulation is defined with $\Sigma[{\CO{2}_O}, \din]\equiv \K^{\{2\}}[{\CO{2}_O}, \din]$, the field-level description in this section uses $\K^{\{2\}}[{\beps{1}_O}, \din]$. 
Hereafter, we usually drop the arguments in $Y$ and $\K^{\{m\}}$ for clarity.
We also drop parameter-independent constants as they are irrelevant for the desired field-level posterior. 
Performing the integral over $\eps$, we have 
\ba
\hatP[\hdgkmax] \big|_{m=1} 
= \int \Del\din \P[\din]  \hat{\N}_{\LL}
& \exp\left[ -\frac12 \v{Y}^T \vhatSigma^{-1} \v{Y} \right] 
\,,
\label{eq:nslike1_orderm1}
\ea
where 
\ba
\hatSigma^{ij}[\{\beps{1}_O\}, \din] &\equiv P_\eps\:  {\K}^{\{2\}ik}{\K}^{\{2\}kj}[\{\beps{1}_O,\din] \vs
\hat{\N}_{\LL}[\{\beps{1}_O\}, \din] &\equiv N_\eps[P_\eps] |\vhatSigma|^{-1/2}
\,.
\label{eq:hatSigma}
\ea
While not immediately obvious, it is straightforward to show that
  $\vhatSigma$ is equivalent to $\v{\Sigma}$ in \refeq{Pgaussgen}. This is
  because for any $O_1$ and $O_2$ in the basis of bias operators, the real-space product
  $O_1 O_2$ is also in the basis. The momentum structure of \refeq{hatSigma} is such that it precisely contains such real-space products (see \refapp{npt} for a related argument at the $n$-point-function level). In addition to \refeq{b1toCO2}, we obtain 
\be
\CO{2}_O = 2 \beps{1}_\unit \beps{1}_O P_\eps\,,
\label{eq:CO_beps}
\ee
for any elementary operator $O$, i.e.~one that cannot be written as a product $O_1 O_2$, and 
\be
\CO{2}_{O_1 O_2} = \left[2 \beps{1}_\unit \beps{1}_{O_1 O_2} + \beps{1}_{O_1} \beps{1}_{O_2}\right] P_\eps \,,
\ee
otherwise. 

We now incorporate higher-order-in-noise terms $m \geq2$. 
For $\eps$ close to Gaussian, we can expand the Dirac delta in terms of its derivatives using its Fourier representation. Considering the field-level Dirac delta for a single mode with index $k$, we have 
\ba
\dirac{1}&\left(Y^k - \K^{\{2\},kj}\eps_{j} - \sum_{m=2}^\infty \K^{\{m+1\},kj_1 \dots j_m}\eps_{j_1} \dots \eps_{j_m} \right) \vs
&=
\int {\rm d}s \exp{\left[\ii s \left(Y^k - \K^{\{2\},kj}\eps_{j} \right)\right]}
\exp{\left[\ii s \left(  - \sum_{m=2}^\infty \K^{\{m+1\},kj_1 \dots j_m}\eps_{j_1} \dots \eps_{j_m}\right)\right]} 
\vs
&=
\sum_{\ell = 0}^\infty \frac{(-1)^\ell}{\ell!} \left[ \sum_{m=2}^\infty \K^{\{m+1\},k j_1 \dots j_m}\eps_{j_1} \dots \eps_{j_m} \right]^\ell \left[\int {\rm d}s (\ii s)^\ell \exp{\left[\ii s \left(Y^k - \K^{\{2\},kj} \eps_{j} \right)\right]}  \right]
\vs
&=\sum_{\ell = 0}^\infty \frac{(-1)^\ell}{\ell!} \left [\sum_{m=2}^\infty \K^{\{m+1\},k j_1 \dots j_m}\eps_{j_1} \dots \eps_{j_m} \frac{\partial}{\partial Y^k} \right]^\ell  \dirac{1}\left(Y^k - \K^{\{2\},kj}\eps_{j}  \right)\,.
\ea
Note that the expansion in $\ell$ here corresponds to an expansion in powers of the \emph{non-Gaussian} stochastic contribution (powers of $\K$ with $m+1\geq 3$ upper indices), while the expansion in $m$ continues to denote the order in the Gaussian noise field $\eps$. The latter expansion is parametrically equivalent to that discussed in \refsec{currentmarginal_gaussian}.

Using this result, the integral over $\eps$ becomes a sum over Gauss integrals with polynomial integrands: 
\ba
\hatP[\hdgkmax] 
&= \int \Del\din \P[\din] \N_\eps \int \Del\eps 
 \exp\left[ -\frac12  P_\eps^{-1} \v{\eps}^T \v{\eps} \right] 
 \vs
 & \hspace{-1cm}\times \sum_{\ell = 0}^\infty \frac{(-1)^\ell}{\ell!} \prod_k \left\{\left [ \sum_{m=2}^\infty\K^{\{m+1\},k j_1 \dots j_m}\eps_{j_1} \dots \eps_{j_m} \frac{\partial}{\partial Y^k} \right]^\ell  \dirac{1}\left(Y^k - \K^{\{2\},kj}\eps_{j}  \right)  \right\}\,,
 \ea
 and, turning the product of component Dirac deltas into a field-level Dirac again, 
\ba
\hatP[\hdgkmax] 
  &= \int \Del\din \P[\din] \N_\eps \int \Del\eps 
 \exp\left[ -\frac12  P_\eps^{-1} \v{\eps}^T \v{\eps} \right] 
 \vs
& \hspace{-1.2cm}\times \sum_{\ell = 0}^\infty \frac{(-1)^\ell}{\ell!} \prod_{\kappa=1}^\ell \left [ \sum_{m_\kappa=2}^\infty\K^{\{m_\kappa+1\},k_\kappa j_{\kappa,1} \dots j_{\kappa,m_\kappa}}\eps_{j_{\kappa,1}} \dots \eps_{j_{\kappa,m_\kappa}} \frac{\partial}{\partial Y^{k_\kappa}} \right]  \dirac{0,\kmax}\left(\v{Y} - \v{\K}^{\{2\}} \v{\eps}  \right) 
\vs
&= \int \Del\din \P[\din] \hat{\N}_{\LL}\sum_{\ell = 0}^\infty \frac{(-1)^\ell}{\ell!} \prod_{\kappa=1}^\ell  \frac{\partial}{\partial Y^{k_\kappa}} 
\vs
&\hspace{-1.3cm}\bigg\{ \sum_{m_\kappa=2}^\infty\K^{\{m_\kappa+1\},k_\kappa j_{\kappa,1} \dots j_{\kappa,m_\kappa}}(\vKtwoinverse \v{Y})_{j_{\kappa,1}} \dots (\vKtwoinverse \v{Y})_{j_{\kappa,m_\kappa}} 
\exp\left[ -\frac12   \v{Y}^T\vhatSigma^{-1} \v{Y} \right] \bigg\}
\vs
&\equiv  \int \Del\din \P[\din] \hat{\N}_{\LL}
 \exp\left[ -\frac12  \v{Y}^T \vhatSigma^{-1} \v{Y} \right]
\left[1 + \sum_{m=1}^\infty \frac1{m!} \bb^{\{m\},i_1\ldots i_m} Y_{i_1}\ldots Y_{i_m}
\right]\,. \label{eq:nslike1m2}
\ea
Notice that contributions with $m_\kappa$ involve the kernels $\K^{\{m_\kappa+1\}}$. In particular, the corrections to the Gaussian likelihood again start at $\K^{\{3\}}$, as expected.
In the last line, we defined the coefficients\footnote{Note that the $\bb^{\{m\}}$ have the same dimensions as the $\tilde\C^{\{m\}}$.}
 \be
 \bb^{\{m\},i_1\ldots i_m} \equiv  \bb^{\{m\},i_1\ldots i_m}[\{\beps{\geq 1}_O\}, \din]
 \ee
as the sum of all the terms containing $m$ powers of $Y$ in this expansion, which are derived in \refapp{nsderivations}. The term linear in $Y$ ($m=1$) can be absorbed by shifting the mean $\mu$ inside $Y$  [see \refeq{Yremap}], and correspondingly the $m=2$ contribution can be absorbed by shifting $\vhatSigma$.  These are examples of higher-order stochastic terms shifting lower-order contributions, a point to which we return below.

\subsection{Discussion}

The correspondence of \refeq{nslike1m2} with \refeq{PexpandFormal}, the main result of the partition function expansion in the current $\v{J}$ derived in \refsec{L} is now clear, which we repeat for convenience:
\ba
\hatP[\hdgkmax] = \int \Del\din \P[\din]&  \N_{\LL}\exp\left[ -\frac12  \v{Y}^T \v{\Sigma}^{-1} \v{Y} \right]
\bigg[ 1 + \sum_{m=3}^\infty \frac1{m!} \tilde{\C}^{\{m\},i_1\ldots i_m}
   Y_{i_1} \cdots Y_{i_m} \bigg]
\,.
\label{eq:PexpandFormalR}
\ea
In particular, at leading order, there is a unique mapping between the $\bb^{\{m\}}$ in \refeq{nslike1m2} and the kernels $\K^{\{m\}}$:
 \ba
 \bb^{\{m\}, j_1\dots j_{m}} \supset\:& m!\, \K^{\{m\},kj'_1 \dots j'_{m-1}}[\{\beps{m-1}_O\},\din] (\vKtwoinverse)_{j'_1 j_1} \dots (\vKtwoinverse)_{j'_{m-1} j_{m-1}} (\vhatSigma^{-1})_{kj_{m}}\,,
\nonumber
 \ea
 which is derived in \refapp{nsderivations}. As we have shown in \refsec{currentmarginal_gaussian} and \refapp{derivations}, the $\K^{\{m\}}$ can in turn be related unambiguously to the $\tilde{\C}^{\{m\}}$ in \refeq{PexpandFormalR}.
 As in the latter case, there are higher-order corrections. In particular, $\bb^{\{m\}}$ also receives contributions from $\K^{\{m+2\}}$ [see \refeq{bbexamples_app}], and from products of kernels $\K$ (in fact, each $\ell$ yields precisely $\ell$ factors of $\K$, and the above relation was derived at $\ell=1$).
 To be explicit, we give just the leading example ($m=3$) here:
 \ba
\bb^{\{3\},j_1j_2j_3} =\:&  \K^{\{3\},kj'_1 j'_2}[\{\beps{2}_O\},\din] (\vKtwoinverse)_{j'_1 j_1} (\vKtwoinverse)_{j'_2 j_2} (\vKtwoinverse)_{kj_3} \,.
\ea

The fact that higher-order contributions in the $\beps{m}_O$ expansion
correct lower-order terms is already clear from \refeq{dgstochL_newrepeat}
(see also \cite{Schmittfull:2018yuk,akitsu/etal:2025}).
Consider the
effective large-scale shot-noise variance, quantified either by the
coefficient of the term quadratic in $Y$ ($m=2$) in \refeq{nslike1m2} or
by computing the power spectrum from \refeq{dgstochL_newrepeat}, which is given by
\be
(\beps{1}_\unit)^2 + (\beps{1}_\delta)^2 \< \delta^2 \> + \ldots
+ 2 (\beps{2}_\unit)^2 + \ldots
\label{eq:Pepseff}
\ee
in the noise-field formulation. In contrast, in the likelihood obtained from the
partition function \refeq{ZofJ}, this contribution is directly given by $\CO{2}_\unit$.\footnote{We ignore loop corrections resulting from the integration over $\din$ such as the $P^{(22)}$ contribution here, focusing solely on the stochastic contributions at a fixed cutoff.}
This mixing of contributions is a downside of the noise-field formulation, since this can create correlations between parameters that negatively affect the sampling efficiency.
Notice that the same higher-order corrections appeared in the formal expansion of the
general likelihood \refeq{PexpandFormalR}, via the additional higher-order contributions to the $\tilde\C^{\{m\}}$ (cf.~the discussion in \refapp{derivations}). Thus, this feature appears generic to expanding the EFT field-level likelihood around the Gaussian approximation. 
Order by order, it can be remedied by a reparametrization, denoting the combination in \refeq{Pepseff} as $\propto P_{\eps,\rm eff}$, with the parameter $P_{\eps,\rm eff}$ replacing $\beps{1}_\unit$, and similarly for the higher-order coefficients.
We emphasize again that no new contributions are introduced by the higher-order corrections; they can all be absorbed by existing, lower-order $\beps{m}_O$.

Recently, Ref.~\cite{Voivodic:2025quw} similarly considered an expansion of the galaxy density including stochasticity as
\ba
\d_g (\vk) =  \sum_{n=0}^\infty\sum_{m=0}^\infty\int_{\substack{\vp_1,\ldots,\vp_n,\\ \vp_1',\ldots,\vp'_m}}
 \diracpi(\vk-\vp_{1\ldots n}-\vp'_{1\ldots m}) \K^{(n,m)}(\vp_1,\ldots \vp_n,\vp'_1,\ldots \vp'_m) \vs \times 
 \din(\vp_1) \cdots \din(\vp_n)  \eps(\vp'_1) \cdots \eps(\vp'_m) \,,
    \label{eq:Rodrigosform}
\ea
for a generic $\K^{(n,m)}$ kernel, and integrated over $\eps$ including terms up to order $\eps^2$ ($m\leq2$), in a very similar way as done here, to obtain non-Gaussian corrections to the likelihood. Note that \refeq{Rodrigosform} likewise only involves a single stochastic field for a single tracer (the multi-tracer generalization is also given there). In our case, the kernels $\K^{(n,m)}$ are explicitly defined in terms of the bias operator kernels as
\be 
\K^{(n,m)}(\vp_1,\ldots \vp_n,\vp'_1,\ldots \vp'_m) \to \sum_{O:\ n(O) \leq n} b^{\{m\}}_O K^{(n)}_O(\vp_1,\ldots \vp_n)\,.
\ee
Crucially, we showed that this set of kernels fully reproduces the partition function \refeq{ZofJ}.

At this point, we should emphasize that we have performed an expansion of the field-level likelihood in the noise-field formulation, \refeq{nslike1}, in order to connect it to the similar expansion of the likelihood derived from the partition function, \refeq{PexpandFormalR}. Crucially, unlike the latter case, \emph{\refeq{nslike1} provides an explicit, normalized probability distribution at arbitrary order in $m$, i.e. fully incorporating non-Gaussian stochasticity.}

Finally, it is worth pointing out another possible route to connecting the noise-field formulation and the general EFT partition function. We have focused so far on the connection at the level of the likelihood. Conversely, one can also obtain a partition function from the noise-field formulation, by integrating over the field $\eps$. \refeq{nsmodel} implies that we can write a partition function of the form
  \ba
  Z_{\rm nf}[\v{J}] \equiv \int \Del\din \P[\din] \int \Del\eps \P[\eps]
  \exp\bigg(J_{k} \bigg[& \dgdet^{-k}[\{\beps{0}_O\},\din] +  \K^{\{2\},kj}[\{\beps{1}_O\},\din] \, \eps_{j} \\
    &  +  \sum_{m=2}^\infty \K^{\{m+1\},kj_1\ldots j_m}[\{\beps{m}_O\},\din] \, \eps_{j_1}\cdots  \eps_{j_m}
   \bigg]\bigg)\,, \nonumber
     \ea
     where the first line contains the deterministic term and the linear coupling to the Gaussian field $\eps$, while the second line contains the nonlinear couplings.
     One can then expand the exponential in these coupling terms. Noting that $\P[\eps]$ is a Gaussian with diagonal (and fixed) covariance given by $P_\eps$, we again obtain
  a Gaussian integral over $\eps$, which leads exactly to the partition function up to order $\v{J}^2$ [\refeq{ZofJ} or \refeq{ZL}], following the matching in \refeq{b1toCO2} and \refeq{CO_beps}. The higher powers in $\eps$ from the expansion of the last line correspondingly lead to the higher powers of $\v{J}$ in the partition function.
  
To summarize, the noise-field approach, where the galaxy density field is described via \refeq{dgstochL_newrepeat}, is equivalent to \refeq{ZofJ} both at the partition function and likelihood levels. However, the resulting likelihood is explicitly normalized, defined everywhere in the joint field space $(\din,\eps)$,  and allows for a robust numerical evaluation. We describe such an implementation and first results in the next section.

\section{Numerical implementation and results}
\label{sec:results}

We mentioned in the last sections that the field-level formulation of the
galaxy density \refeq{dgstochL_newrepeat} is well suited for practical inference applications.
We now discuss the implementation in \leftfield\ and first results.
We show field-level inference results of $\sigma_8$
from dark matter halos in the rest frame, precisely the case considered
in \cite{FBISBI,beyond2pt}, but now including both non-Gaussian stochasticity
and the coupling between stochasticity and density perturbations, via the
bias terms $\beps{1}_\d$ and $\beps{2}_\unit$. Moreover, we include the leading
higher-derivative stochastic correction via a term $\beps{1}_{\lapl} \lapl\eps$. 

Field-level inference proceeds by numerically sampling from
$\P[\{\theta\},\, \{ \beps{m}_O \} \ |\ \hdgkmax]$, the posterior, where we again
denote the observed data by $\hdgkmax$. The forward model described by
\refeqs{dgstochL_newrepeat}{priordineps} corresponds to
using \refeq{nslike1} instead of \refeq{hatPdef}:
\ba \label{eq:nspost}
\P\left[\{\theta\},\, \{ \beps{m}_O \} \ |\ \hdgkmax\right] &\propto
\int \Del\din \P[\din|\{\theta\}]\int \Del\eps \P[\eps|c_\eps]
 \vs
 &\quad\times \dirac{0,\kmax}\left[\hdg(\vk) - \sum_{m=0}^\infty \sum_{\unit,O}  \beps{m}_O \left(\eps^m O\right)(\vk) \right] \vs
 &\quad\times \P(\{\beps{m}_O\},\ \{\theta\})\,.
\ea
Notice that the likelihood is replaced by a Dirac delta, since all stochastic
contributions are explicitly accounted for by the terms involving $\eps$.
The Dirac delta however only considers the data up to the momentum cut $\kmax$,
as indicated by the superscript. We employ a Gaussian prior with diagonal
covariance on the grid where $\eps$ is discretized (see below),
\be
\P(\eps|c_\eps) \propto \prod_{\vx} \N(\eps(\vx)|0, c_\eps^2)\,,
\ee
which thus multiplies all stochastic contributions.\footnote{The parameter $c_\eps$ is related to the power spectrum $P_\eps$ introduced in the previous section via $P_\eps = (L/N_g)^3 c_\eps^2$, where $L$ is the box and $N_g$ the grid size.} We fix $\beps{1}_\unit=1$,
as it is degenerate with $c_\eps$ (conversely, one could choose $c_\eps=1$ fixed and leave $\beps{1}_\unit$ free). 
For field-level inference, the Dirac likelihood can only be approximated
numerically, and in the actual implementation we replace it with a Gaussian
with fixed variance $\s_0^2$: 
\be
\dirac{0,\kmax}\left[X(\vk)  \right]
  \longrightarrow \exp\left[ -\frac12\sum_{\vk\neq \v{0}}^{\kmax}\left( \frac{|X(\vk)|^2}{\sigma_0^2} + \ln\sigma_0^2\right)\right]\,,
  \label{eq:nslike}
\ee
which asymptotes to the desired Dirac distribution in the limit $\sigma_0\to 0$.
The choice of $\sigma_0$ corresponds to a tradeoff between accuracy of the
likelihood approximation (smaller $\sigma_0$) and numerical efficiency
(larger $\sigma_0$). We have tested that the precise value of $\sigma_0$
has limited significance if it is much smaller than the physical noise
contribution described by $c_\eps$. In practice,
we choose $\sigma_0^2$ to be $\lesssim 0.25$ of the noise variance given by
$c_\eps^2$.
On the other hand, sampling mock datasets from the posterior \refeq{nspost}, as nedeed for simulation-based inference for example, is no problem, and there is no need to introduce a finite $\sigma_0$.

The Gaussian likelihood \refeq{nslike} further allows for analytical marginalization over all $\beps{m}_O$ (cf. \cite{elsner/etal}). We employ this marginalization for all bias
coefficients, listed here for completeness (see \cite{FBISBI} for the definitions)
\be
\left\{\beps{0}_O :O\in\left[\d,\d^2,K^2,\d^3,K^3,\d K^2,\Otd,\nabla^2\d\right]
  \right\},\  \beps{1}_{\lapl},\  \beps{1}_\d ,\  \beps{2}_\unit\,,
    \ee
    except for $\beps{0}_\delta\equiv b_1$ and $\beps{1}_\unit$, the latter of which is fixed.\footnote{Elsewhere, the notation $b_{\eps\delta}\equiv\beps{1}_\d$, $b_{\eps^2}\equiv\beps{2}_\unit$, $b_{\lapl\eps}\equiv \beps{1}_{\lapl}$ is also used.}
The deterministic bias expansion is chosen to match that of \cite{FBISBI}, so that only the stochastic part of the model changes. Here, we decide to keep the leading stochastic terms that appear in the galaxy power spectrum and bispectrum, as well as the subleading stochastic term in the power spectrum ($\beps{1}_{\lapl}$).
The priors on $\sigma_8$ and the bias coefficients are chosen as in \cite{FBISBI},
  with wide priors for the additional stochastic coefficients,
  \ba
  \P(c_\eps) &= \mathcal{U}(0.05, 0.5); \quad
  \P(\beps{1}_{\lapl}\  [(\Mpch)^2]) = \N(0, 5^2 ); \vs
\P(\beps{1}_\delta) &= \N(0, (0.5)^2); \quad
\P(\beps{2}_\unit) = \N(0, (0.2)^2)\,.
  \ea
The restriction of the noise expansion to second order is justified by the low noise of the sample considered, i.e.~the fact that the effective noise variance at $k_{\rm max}$, $\sim 0.014$ for the empirically found value of $c_\eps =0.2$ (see below), is much less than $\< [\d^{(1)}_{\kmax}]^2 \> \sim 0.14$ [cf. the discussion after \refeq{nslike1}].

To sample from the posterior \refeq{nspost}, we employ joint Hamiltonian Monte Carlo (HMC) sampling
of the two fields $\{ \hat s, \eps \}$, which both have unit Gaussian
priors, where $\hat s$ is related to $\din$ via
\be
\din(\vk) \propto \sqrt{\Plin(k)} \hat s(\vk),
\label{eq:shat}
\ee
and the normalization depends on the grid size (see e.g.~\cite{FBISBI}).
Both $\hat s$ and $\eps$ are filtered at a scale $\Lambda > \kmax$ and
represented on grids that are sized appropriately to have Nyquist frequency
just above $\Lambda$.
Here, we choose $\Lambda=0.14\iMpch$ and $\kmax=0.12\iMpch$, the higher
cutoff values considered in \cite{FBISBI}. 
We employ a block-diagonal mass matrix, consisting of dense $2\times 2$ blocks
for each pair $\{ \hat s(\vk), \eps(\vk) \}_{\vk}$, with components derived
from a linear forward model, the only case where the posterior can be computed
analytically \cite{Kostic:2022vok}. The HMC sampling steps are interleaved with slice
sampling steps for the parameters $\{\beps{0}_\d, c_\eps, \sigma_8\}$ in a block-sampling fashion.

Given a proposal for $\{ \hat s(\vk), \eps(\vk) \}_{\vk},\  \beps{0}_\d, c_\eps, \sigma_8\}$, the forward model and field-level likelihood evaluation proceed as follows:
\begin{enumerate}
\item $\din$ is computed via \refeq{shat}, and the grids for $\din,\eps$ are zero-padded to avoid aliasing to modes below $\kmax$ in the computation of the nonlinear operators.
\item Lagrangian perturbation theory and bias expansions are performed to construct the Eulerian operators $O$. The details can be found in \cite{stadler:2024a}. In the present case, we employ second-order LPT (2LPT) and a third-order Eulerian bias expansion. This was the matter/bias model employed in \cite{FBISBI}.
\item In parallel, the fields $\eps^m$ are constructed (here we restrict to $m \leq 2$) and the $\eps(\vx) O(\vx)$ are constructed in Eulerian space. Zero-padding is again performed as necessary to avoid aliasing to modes below $\kmax$.
\item Finally, the likelihood \refeq{nslike} is evaluated, where $\sigma_0$ is fixed, i.e. not varied in the inference.
\end{enumerate}

Conceptually, this sampling approach is straightforward, can easily be
extended to any desired order in the expansion \refeq{dgstochL_newrepeat},
and also allows for the accurate incorporation of other physical effects
that are beyond the scope of this paper, such as redshift-space distortions
\cite{Cabass:2020jqo,stadler:2024b}. The major drawback is the need to sample
two correlated fields, and thus not only doubling the dimensionality of the inference problem, but also adding significant correlations.
In the present case, we have $2\times 90^3 \approx 1.5$~million free
parameters. The increased dimensionality and correlations lead to less efficient exploration
of the posterior space, i.e. longer correlations between samples.

\begin{figure}[t]
	\centering
	\includegraphics[width = 0.49\textwidth, trim=0.7cm 0cm 0.5cm 0cm, clip=true]{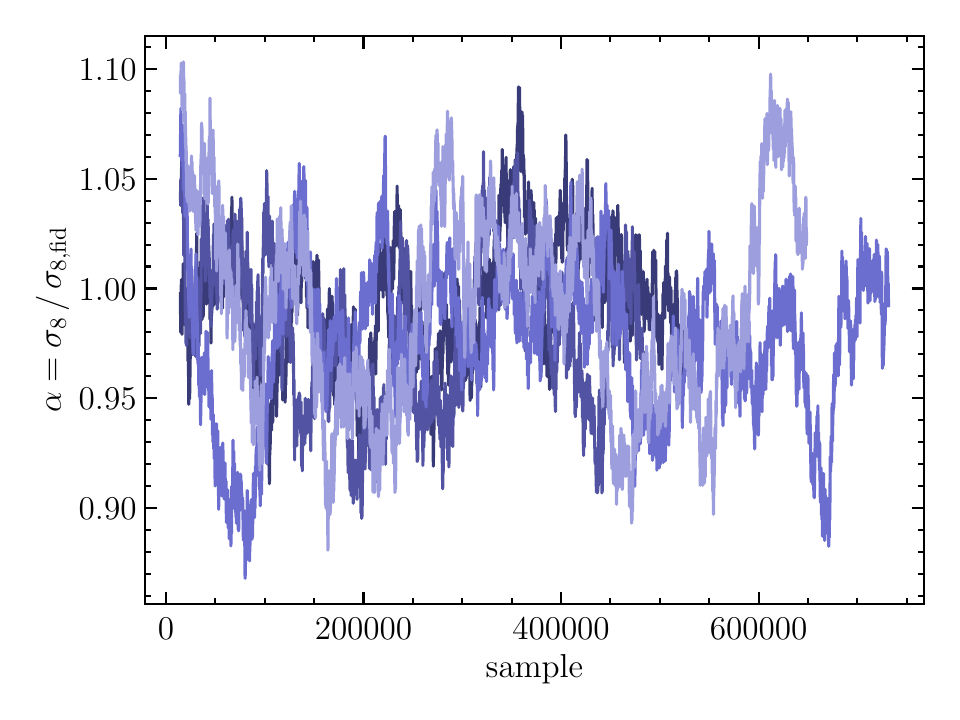}
	\includegraphics[width = 0.49\textwidth, trim=0.7cm 0cm 0.5cm 0cm, clip=true]{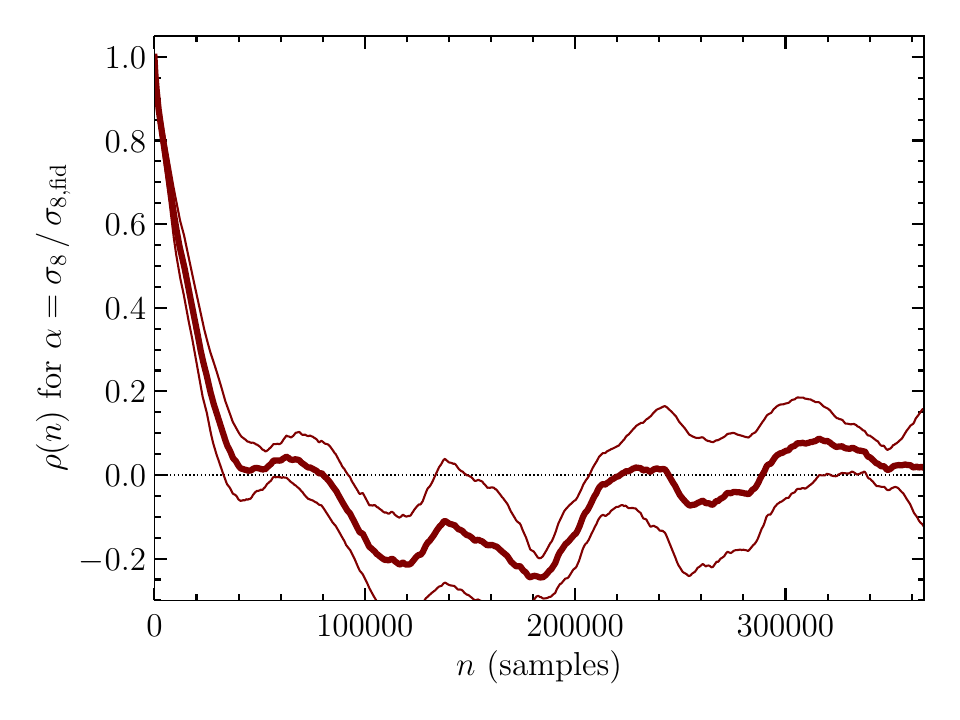}
	\caption{Trace plot (left panel) and normalized autocorrelation function $\rho(n)=\xi(n)/\xi(0)$ for the parameter $\alpha \equiv \sigma_8/\sigma_{8,\rm fid}$ in four independent FLI chains for the ``SNG'' rest-frame dark matter halo sample from \cite{FBISBI}. The autocorrelation for each chain is estimated after removing a burn-in phase of 100,000 samples. The thick line in the right panel shows the mean autocorrelation across the four chains, while the thin lines indicate the error on the mean estimated from the sample variance across correlation functions.}
	\label{fig:alphatrace}
\end{figure}

We first consider the parameter $\alpha \equiv \sigma_8/\sigma_{8,\rm fid}$.
\reffig{alphatrace} shows parameter traces from four independent
sampling chains (left panel) and the autocorrelation function averaged over
the four samples after removing burn-in (right panel).
Each chain was initialized with different values of the parameters
  $\{\beps{0}_\d, \beps{1}_\d, c_\eps, \sigma_8\}$, with the other higher-order parameters initially set to zero.
Estimating the
correlation length $\tau$ as the first crossing of $\rho(\tau) = 0.1$,
we obtain $\tau \simeq 32,600$, with a range of $26,900 - 40,300$ estimated
from cross-chain variance (reducing the threshold further from 0.1 does not affect
$\tau$ significantly, as can be gleaned from the figure). 
Clearly, a very large Monte Carlo sample size is required to obtain
converged statistics (in this case $\sim 2$~million after removing $400,000$ burn-in samples in total).\footnote{The total compute time amounts to $\sim 6,200$ CPU hours on Intel Xeon 8480+ CPUs @ 2 GHz.}
The total effective sample size is estimated
to be 61 (range $49 - 74$), and the Gelman-Rubin statistic is $R(\alpha)=1.06$. 
Given this effective sample size, it is justified to report the mean and 68\% CL error bar for $\alpha$:
\be
\alpha = 0.984 \pm 0.035\,.
\ee
\emph{This result corresponds to the first field-level cosmology inference that
uses the proper EFT-based, non-Gaussian noise model.} The error bar is very mildly increased over
that reported for Gaussian noise in \cite{FBISBI}, who obtained $\alpha=1.013\pm 0.033$, while the posterior mean is consistent with the latter within $< 1\sigma$.
It is worth emphasizing that this is a high number density, i.e.~low-noise, halo sample with $\bar n \simeq 1.3\cdot10^{-3}(\Mpch)^{-3}$. 
This explains why the detailed noise model does not affect the inference of the power spectrum amplitude $\sigma_8$ significantly.\footnote{Note that the recent Ref.~\cite{akitsu/etal:2025} considered mock datasets with much higher noise amplitude.}
While this finding could still be considered preliminary given the limited
effective sample size (the results in \cite{FBISBI} were based on an effective sample
size greater than 100), it indicates that a non-Gaussian noise model does not
substantially affect field-level inference error bars on $\sigma_8$,
at least for high-number-density samples.

\begin{figure}[t]
	\centering
	\includegraphics[width = 0.49\textwidth, trim=0.7cm 0cm 0.5cm 0cm, clip=true]{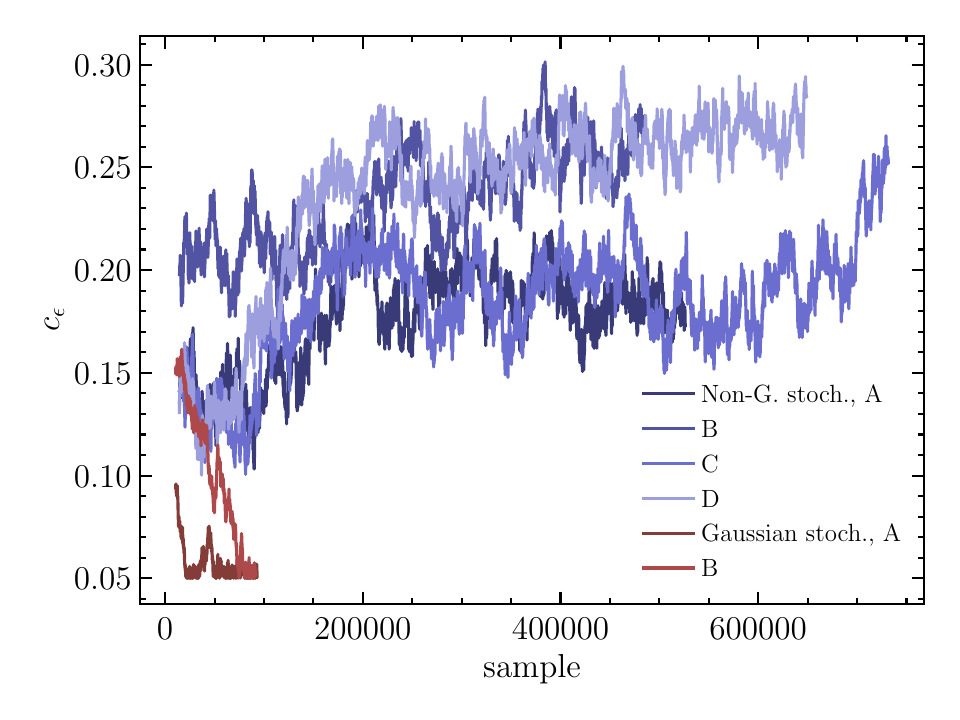}
	\includegraphics[width = 0.49\textwidth, trim=0.7cm 0cm 0.5cm 0cm, clip=true]{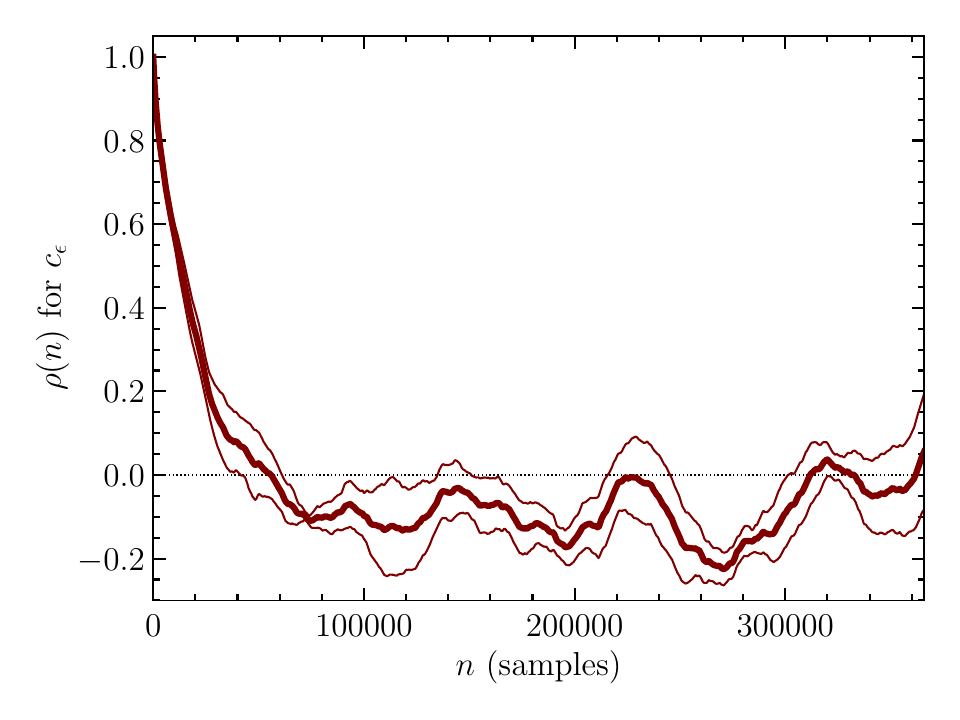}
	\caption{Trace plot (left panel) and normalized autocorrelation function $\rho(n)=\xi(n)/\xi(0)$ for the noise-amplitude parameter $c_\eps$ for the same chains as in \reffig{alphatrace} (blue shades, labeled A--D). In the trace plot we also show the results of two chains using the same noise-field formulation but with non-Gaussian and density-dependent noise turned off ($\beps{1}_\d = 0 = \beps{2}_\unit$; maroon shades). These latter chains show that $c_\eps$ drifts to its lower limit $0.05$, a trend previously found for Gaussian likelihoods in \cite{nguyen/etal:2020,FBISBI,akitsu/etal:2025}.}
	\label{fig:cepstrace}
\end{figure}

We now turn to the inference of the noise amplitude parameter $c_\eps$.
Ref.~\cite{nguyen/etal:2020}, and more recently \cite{FBISBI,akitsu/etal:2025}, pointed out
that the noise amplitude drifts
to unphysically small values in FLI inferences applied to nonlinear tracers such as halos \cite{nguyen/etal:2020,FBISBI}, HOD-based catalogs \cite{beyond2pt} or mock data with non-Gaussian noise \cite{akitsu/etal:2025}, when assuming Gaussian stochasticity in the inference.
The term ``sigma collapse'' was coined for this phenomenon \cite{nguyen/etal:2020}, given that the variance in the Gaussian field-level likelihood is usually denoted as $\sigma^2$. 
We can test for this phenomenon in the noise-field formulation as well, by
performing an inference where $\beps{1}_\delta$ and $\beps{2}_\unit$ are fixed to zero. 
The resulting parameter traces for $c_\eps$ are shown as maroon lines in the left panel of \reffig{cepstrace}, which clearly reproduce ``sigma collapse,'' i.e. the drift of the noise
amplitude $c_\eps$ to small values (0.05 being the arbitrary lower bound imposed
on the parameter here).

The blue lines in the left panel of \reffig{cepstrace} instead show the traces for the same chains as in \reffig{alphatrace}, i.e.~allowing for density-dependent and non-Gaussian stochasticity. Evidently, \emph{``sigma collapse'' does
not occur in FLI chains that consistently incorporate non-Gaussian stochasticity.} Instead,
$c_\eps$ remains at a physically expected level; the pure Poisson-noise
expectation, $P_\eps = 1/\bar n$, for this halo catalog corresponds to $c_\eps^{\rm Poisson} = 0.15$.
Note that one expects a somewhat larger effective noise from integrating out modes above
the cutoff $\L$ for a nonlinearly biased tracer \cite{Kostic:2022vok,Rubira:2024tea}. Further deviation from Poisson statistics could also be present due to halo-exclusion effects \cite{Baldauf:2013hka}.

\begin{figure}[t]
	\centering
	\includegraphics[width = 0.49\textwidth, trim=0.45cm 0cm 0.5cm 0cm, clip=true]{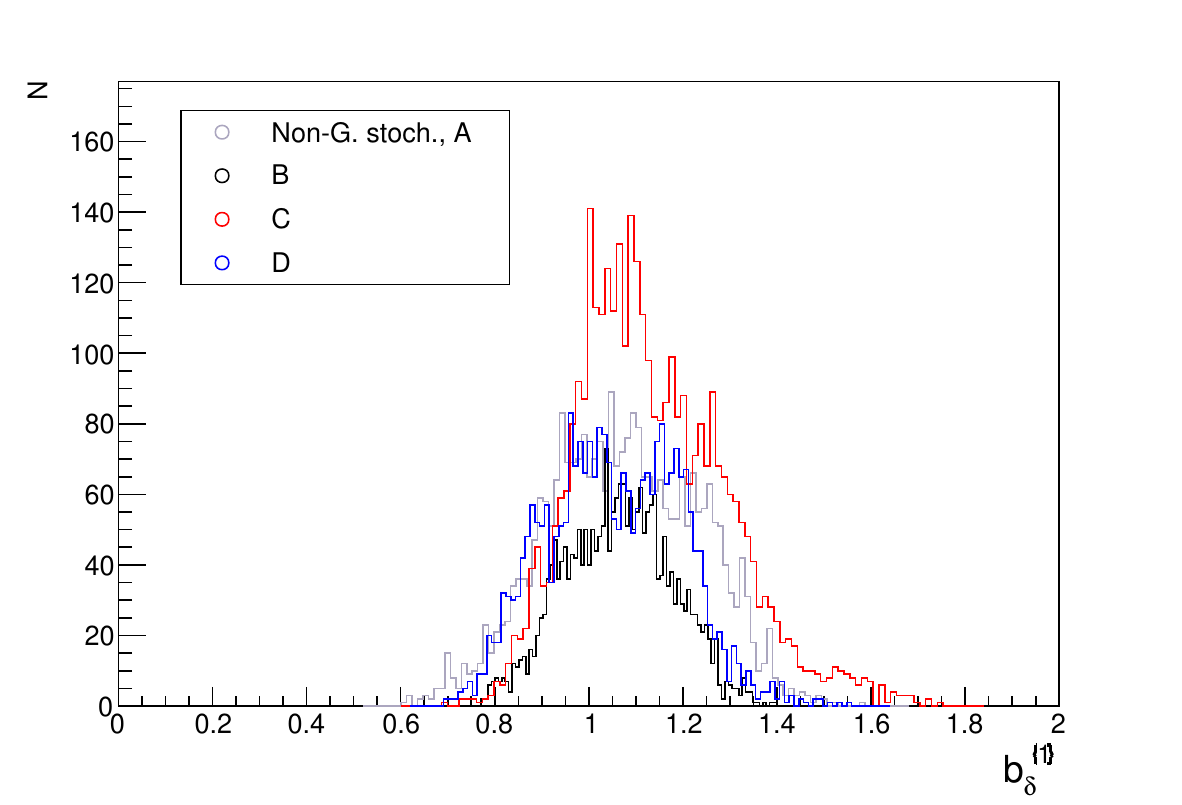}
	\includegraphics[width = 0.49\textwidth, trim=0.45cm 0cm 0.5cm 0cm, clip=true]{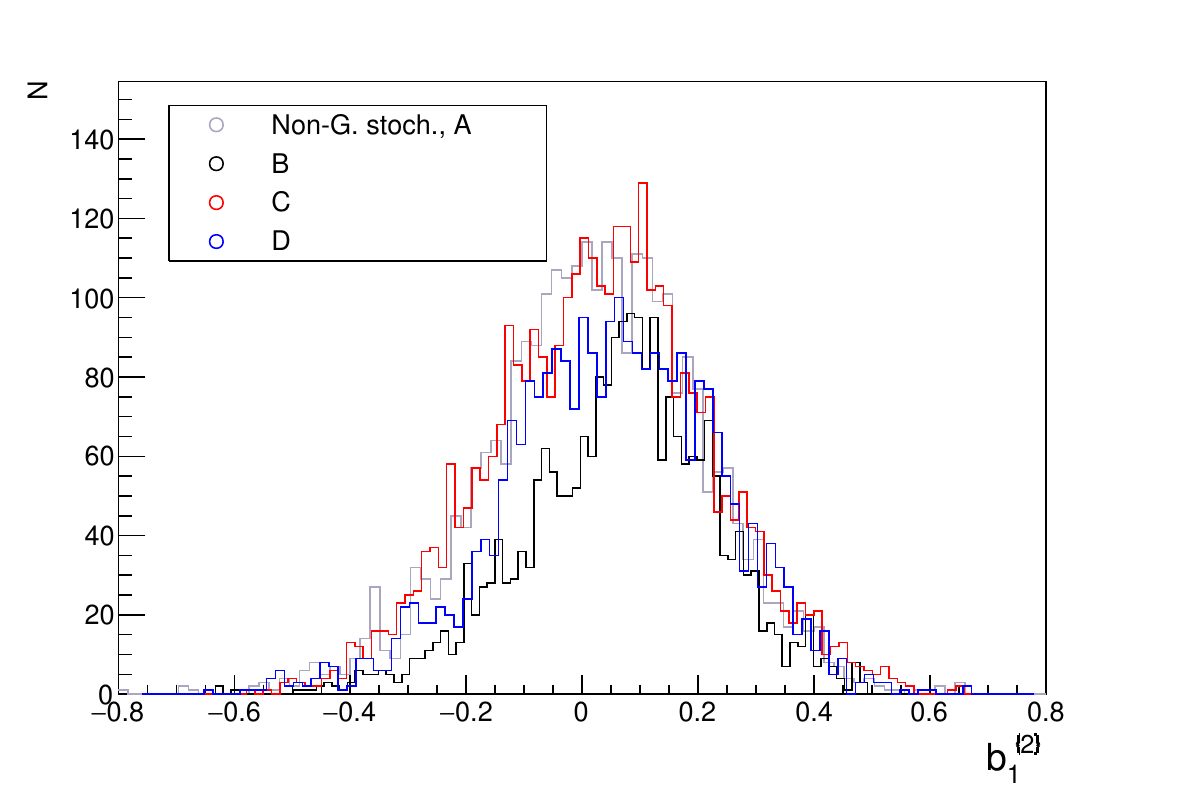}
	\caption{Posterior histograms for $\beps{1}_\d$ (left) and $\beps{2}_\unit$ (right) inferred from the four FLI chains with non-Gaussian stochasticity. The former is significantly nonzero and well constrained, while the latter is poorly constrained by the data (posterior is close to consistent with the prior) for the halo sample considered.
        }
	\label{fig:bepshist}
\end{figure}

Nevertheless, sampling $c_\eps$ remains challenging due to the long
correlations within samples.
A residual dependence on the starting value of $c_\eps$ is still clearly visible in the parameter traces.  Hence, we do not quote any posterior mean or error bar here.
While the overdispersion in $c_\eps$ fortunately does not affect the posterior for $\alpha$
strongly (as evidenced by the Gelman-Rubin statistic for $\alpha$ reported above), proper robust cosmology posteriors require converged posteriors in all parameters. Future work must thus aim at improving the sampling efficiency.
This could involve an improved mass matrix, different marginalization schemes, and/or joint HMC sampling of parameters and fields.

Finally, we present posterior histograms for the non-Gaussian stochastic coefficients in \reffig{bepshist}. These were obtained in post-processing by sampling from the conditional posteriors $\{\P(\beps{m}_O | \hdgkmax, \hat{s}_k, \eps_k, ...)\}_{k=1}^{N_{\rm sample}}$ for a subsample $k$ (roughly every 100th) of the MCMC chains.

We have also obtained preliminary results for FLI using the same forward model and bias/stochastic expansion on a halo sample with mass range $\log_{10} M/ h^{-1} M_{\odot} \in [13, 13.5]$ at $z=0.5$, with a much lower mean number density of $\bar n = 3.2\times 10^{-4} (h^{-1} {\rm Mpc})^{-3}$. We have found that $c_\eps$ remains stable also for this case. However, this halo sample shows evidence of prior volume effects, specifically, the $\alpha$ posterior depends on the prior volume in bias parameter space; while it is significantly biased for the fiducial priors, reducing the prior variance in second- and third-order bias parameters from $1^2$ to $0.3^2$ leads to an unbiased posterior (within $1\sigma$). In addition to exploring appropriate bias priors and reparametrizations to reduce prior volume effects, one should also verify whether third-order stochastic contributions become relevant for this higher-noise sample, since the variance of $\eps$ may be comparable to the variance of density perturbations. Since these issues are beyond the scope of this paper, we refrain from showing explicit results for this sample, and postpone this to future work.

\section{Conclusions}
\label{sec:concl}

In this paper, we have built on the general EFT partition function for
galaxy clustering from \cite{Rubira:2024tea}, \refeq{ZofJ}, to investigate how stochastic
contributions to galaxy clustering can be described at the field level.
Our results are at two levels.
First, we derive the general expression for the field-level likelihood
in the EFT, which is given by the functional Fourier transform of the
partition function (\refsec{posterior_from_Z}). This Fourier transform cannot be computed in closed form. However, one
can expand around the Gaussian limit of stochasticity (\refsec{currentmarginal_gaussian}), which is a valid expansion within the perturbative EFT context.
Second, we establish that the reduced model with a single Gaussian field,
\refeq{dgstochL_newintro}, is a sufficient description of galaxy stochasticity
within the EFT, by deriving the same likelihood in this formulation (\refsec{NS}).
This description is significantly more restrictive than the non-minimal model
in \refeq{dgstochL_oldintro} which has so far been assumed as the standard.
\refapp{npt} compares the two approaches at the $n$-point function level.

The formal likelihood obtained from the EFT partition function has two major drawbacks: it is not normalized, showing
unphysical behavior in the tails (i.e. negative probability densities), as is
common for Edgeworth-like expansions of probability distributions around a Gaussian; and it involves field-level matrix inversions which are intractible in practice (see also \refapp{reallikelihood}). 
Instead, the \emph{noise-field formulation} of the EFT likelihood presented in \refsec{NS} avoids unphysical probabilities and is suited for numerical sampling, while
at the same time capturing the full EFT likelihood, order by order.

Finally, \refsec{results} presents first results of field-level cosmology inference
using this noise-field formulation. The downside of this approach is that the dimensionality of
parameters to be sampled is doubled, yielding a slower exploration
of the parameter space. Nevertheless, we show converged results for $\sigma_8$
(\reffig{alphatrace}) jointly inferred with bias and stochastic parameters from a dark matter
halo sample in real space. 
This represents the first time that the full EFT model for galaxy bias
and stochasticity has been employed in field-level inference.
Moreover, the non-Gaussian noise formulation leads to a stable inferred noise amplitude at a physically expected value, and thus solves the problem of the ``collapse'' of the noise variance observed for the Gaussian-noise case (\reffig{cepstrace}).

In the future, it will be imperative to explore more efficient sampling techniques,
perhaps including more refined analytical likelihoods that correctly capture the stochastic
mode couplings which the existing real-space likelihood unfortunately does not (\refapp{reallikelihood}). 
It would also be interesting and important to generalize the results of our work to predict the  stochastic contributions to the galaxy velocity field, which are necessary for redshift-space distortions.

\acknowledgments

The authors acknowledge Minh Nguyen and Zvonimir Vlah for useful conversations, and Francesco Conteddu and Mathias Garny for comments on the draft.

\appendix

\section{Comparison of standard (non-minimal) and minimal stochastic contributions for $n$-point functions}
\label{app:npt}

We here compare the predictions for stochastic contributions to galaxy $n$-point functions predicted by \refeq{dgstochL_oldintro} on the one hand, and \refeq{dgstochL_newintro} on the other.
Consider a ``Gaussian stochastic'' contribution to the galaxy $n$-point correlation function in real space,
\be
\< \d_g(\vx_1) \cdots \d_g(\vx_n) \> ,
\ee
defined as involving precisely two powers of stochastic fields. We choose to
work in real space here, in order to make the local products of operators
simple to express. In the
non-minimal formulation [\refeq{dgstochL_oldintro}], such a contribution can be written as
\ba
&\< \epsold_O (\vx_1) O(\vx_1) \epsold_{O'}(\vx_2) O'(\vx_2) O_3(\vx_3) \cdots O_n(\vx_n)\> \vs
& =\< \epsold_O (\vx_1) \epsold_{O'}(\vx_2) \>\: \< O(\vx_1) O'(\vx_2) O_3(\vx_3) \cdots O_n(\vx_n)\> \vs
& = P_{\eps_O \eps_{O'}} \diracc(\vx_1-\vx_2)  \< O(\vx_1) O'(\vx_1) O_3(\vx_3) \cdots O_n(\vx_n)\>\,,
\label{eq:stoch2npt-old}
\ea
where $O, O'$ could be either the unit operator or a nontrivial bias operator, while $O_3,\ldots O_n$ stand for arbitrary bias operators (otherwise, we trivially reduce to a lower order $n$-point function). We have also introduced
\be
\< \epsold_O(\vk) \epsold_{O'}(\vk') \>' = P_{\eps_O \eps_{O'}}\,.
\label{eq:PepsOold}
\ee
A $k^2$-dependence capturing higher-derivative stochastic terms following \refeq{hd_exp_noise} can also be included here, corresponding to derivatives acting on the Dirac delta in \refeq{stoch2npt-old}.

On the other hand, \refeq{dgstochL_newintro} predicts\footnote{At leading order; as discussed in \refsec{NS}, these contributions are corrected by higher-order terms, which however have the same structure and can be absorbed in redefined $\beps{1}_O$.}
\ba
& \beps{1}_O \beps{1}_{O'} P_\eps  \diracc(\vx_1-\vx_2)  \< O(\vx_1) O'(\vx_1) O_3(\vx_3) \cdots O_n(\vx_n)\>\,,
\label{eq:stoch2npt-new}
\ea
where $P_\eps$ is introduced in \refeq{priordineps}. 
Clearly, the two expressions are equivalent in terms of structure. To
understand the apparent reduction in coefficients in the two formulations,
consider a pair of operators, such as $\unit$ and $O$. This
pair is characterized by three coefficients in the non-minimal formulation,
$P_{\eps_\unit\eps_\unit}, P_{\eps_\unit\eps_O}, P_{\eps_O \eps_O}$. Indeed,
three coefficients are needed to describe the covariance of two fields.
On other hand, the same pair is only controlled by two coefficients in
\refeq{stoch2npt-new}, $\beps{1}_\unit, \beps{1}_O$. However, the contribution
to \refeq{stoch2npt-old} that comes with $P_{\eps_O \eps_O}$ is precisely degenerate with that controlled by $P_{\eps_\unit \eps_{O  O}}$, which has an anologous
coefficient $\beps{1}_{O  O}$ in \refeq{stoch2npt-new}. This degeneracy
is due to (1) the locality of the stochastic process; and (2) the completeness
of the bias expansion, which for any $O_1, O_2$ in the basis also includes
$O_1  O_2$. In a practical analysis, one would thus eliminate either $P_{\eps_O \eps_O}$ or $P_{\eps_\unit \eps_{O  O}}$.
Thanks to these facts, \refeq{stoch2npt-new} can capture the full set of non-degenerate contributions in \refeq{stoch2npt-old}.

To illustrate how this reasoning continues to higher order, consider a contribution with three powers of stochastic fields.
\refeq{dgstochL_oldintro} yields
\ba
&\< \epsold_O (\vx_1) O(\vx_1) \epsold_{O'}(\vx_2) O'(\vx_2)\epsold_{O''}(\vx_3) O''(\vx_3) O_4(\vx_4) \cdots O_n(\vx_n)\> \vs
& =\< \epsold_O (\vx_1) \epsold_{O'}(\vx_2) \epsold_{O''}(\vx_3) \>\: \< O(\vx_1) O'(\vx_2) O''(\vx_3) O_4(\vx_4) \cdots O_n(\vx_n)\> \vs
& = B_{\eps_O \eps_{O'}\eps_{O''}} \diracc(\vx_1-\vx_2) \diracc(\vx_2-\vx_3)  \< O(\vx_1) O'(\vx_1) O''(\vx_1) O_4(\vx_4) \cdots O_n(\vx_n)\>\,,
\label{eq:stoch3npt-old}
\ea
where $B_{\eps_O \eps_{O'}\eps_{O''}}$ is defined analogously to \refeq{PepsOold}, 
while \refeq{dgstochL_newintro} predicts
\ba
& 2 \left[ \beps{2}_O \beps{1}_{O'} \beps{1}_{O''} + 2\:{\rm perm.}\right] (P_\eps)^2 
\diracc(\vx_1-\vx_2) \diracc(\vx_2-\vx_3) \vs
& \qquad\times\< O(\vx_1) O'(\vx_1) O''(\vx_1) O_4(\vx_4) \cdots O_n(\vx_n)\>\,.
\label{eq:stoch3npt-new}
\ea
Again, both have the same structure. Moreover, since $B_{\eps_O \eps_{O'}\eps_{O''}}$ is totally symmetric (as the configuration and scale dependence of $B_{\eps_O \eps_{O'}\eps_{O''}}$ is trivial, it is invariant under interchange of any of the fields $\eps_O$), and its contribution is degenerate with
that of $B_{\eps_\unit \eps_{O  O'} \eps_{O''}}$ as well as other corresponding
terms, \refeq{stoch3npt-new} has sufficient flexibility to
capture the non-degenerate contributions to \refeq{stoch3npt-old}.

\section{Issues with the real-space formulation of the likelihood}
\label{app:reallikelihood}

The real-space likelihood formulation introduced in \cite{Cabass:2020nwf} in principle
offers a neat way of incorporating the stochastic terms $\CO{2}_O$ or
equivalently $\beps{1}_O$, i.e. the leading coupling between stochasticity
and long-wavelength modes while keeping the stochasticity Gaussian.
This is precisely the case studied in \refsec{currentmarginal_m2}. The likelihood
in \refeq{Pgaussgen} cannot be simply computed in Fourier space, as the covariance
$\Sigma$ is dense. However, in real space we can, at least naively, write
\refeq{KKdef} as\footnote{The $\COreal{2}_{\unit, O}$ appearing here are normalized differently than those in \refeq{Pgaussgen}, since they refer to real-space fields and the reduced grid $N_g$.}
\be
\Sigma_{\vx,\vy} = \left[\COreal{2}_\unit + \sum_O \COreal{2}_O O[\din](\vx)\right]
\diracc(\vx-\vy)\,,
\ee
since the coupling between stochasticity and the $O[\din]$ is local in real space. This is a diagonal covariance matrix which can be trivially inverted in real space.
However, one still needs to implement the sharp-$\kmax$ cut
in the likelihood in \refeq{Pgaussgen}. For this, Refs.~\cite{Cabass:2020nwf,Schmidt:2020tao} proposed to
perform a grid reduction in Fourier space, effectively restricting
the components $i=1,2,3$ of all represented wavenumbers to $|k_i| \leq k_{\rm Ny} = N_g \pi/L$, where $k_{\rm Ny}$ is the Nyquist frequency, $N_g$ the grid resolution, and $L$ the size of the box.
By appropriately choosing $N_g$, one can ensure that $|k_i| \leq \kmax$.
The real-space likelihood is then given by \cite{Cabass:2020nwf,Schmidt:2020tao}
\ba
\LL_{\rm real}(\hdg | \{b_O\}, \{\COreal{2}_O\}, \din) &\propto \exp\left[ -\frac12\sum_{\vx}^{N_g^3}\left( \frac{\left(\hdg(\vx) - \dgdet[\din, \{b_O\}](\vx)\right)^2}{\sigma^2(\vx)} + \ln\sigma^2(\vx)\right)\right]\,, \vs
\textrm{with} \quad \sigma^2(\vx) &\equiv \sigma^2[\din,\{\COreal{2}_O\}](\vx) = \COreal{2}_\unit + \sum_O \COreal{2}_O O[\din](\vx)\,,
\label{eq:reallike}
\ea
where we have explicitly indicated the grid size in the real-space sum,
and both $\hdg$ and $\dgdet$ are reduced to this grid in Fourier space
before evaluating the likelihood (in an actual implementation, one would also ensure that $\sigma^2(\vx)$ is positive definite, see \cite{Schmidt:2020tao}).

We now want to show why \refeq{reallike} does not correctly implement the coupling of stochasticity to long-wavelength modes. In order to do this, we
  consider the generating process corresponding to \refeq{reallike}, which makes
  the issue more explicit. Hence, imagine generating
a mock dataset, i.e. tracer field from the real-space likelihood. For simplicity,
we consider a fixed deterministic prediction $\dgdet(\vx)$, since this field
does not play a role in this discussion. Clearly, such
a mock dataset will again be represented on a grid of size $N_g$, and can
be written as
\ba
\hat\delta_{g,\rm real}(\vx) &= \dgdet(\vx) + \eps_{\rm real}(\vx) \quad \textrm{with}\vs
\eps_{\rm real}(\vx) &\sim \mathcal{N}(0, \sigma^2(\vx)) = \left[\COreal{2}_\unit + \sum_O \COreal{2}_O O(\vx)\right]^{1/2} \eps_{N_g}(\vx)\,,
\label{eq:epsreal}
\ea
where $\eps_{N_g}(\vx) \sim \mathcal{N}(0,1)$ is a unit Gaussian random
field generated on the grid $N_g$ (and again, all of $\hat\d_{g,\rm real}, \dgdet, O$ are represented on the same grid).

The last line in \refeq{epsreal} bears out the issue. Note first that both $\eps_{N_g}$ and
$O$ have Fourier-support up to the Nyquist frequency of the grid,
$k_{\rm Ny} = \kmax$. This is necessary in order to have the correct mean-field prediction $\dgdet$ and leading noise contribution up to $\kmax$. On the other hand, \refeq{epsreal} 
multiplies
the fields $\eps_{N_g}$ and $O$ in real space, thus exciting modes up to
$2 k_{\rm Ny}$ in Fourier space (it is sufficient to expand the square-root
in \refeq{epsreal} to linear order in perturbations to see this).
These modes cannot be represented directly on the
grid of size $N_g$, and are instead aliased to lower-$k$ modes. In fact,
\emph{all} Fourier modes on the likelihood grid are polluted by aliasing.
This aliasing is unphysical, since the grid size is related to $\kmax$, a scale which
has no physical significance for the tracer or forward model.
On the other hand, increasing the likelihood grid size while maintaining the $\kmax$ cut is not possible.

It thus appears impossible to represent the correct mode-coupling structure
of the stochastic terms $\propto \CO{2}_O$ in \refeq{Pgaussgen} in a
closed-form real-space likelihood, while at
the same time having a sharp filter on the data at some $\kmax < \L$.

\section{Derivation of coefficients $\C$ and $\tilde{\C}$}
\label{app:derivations}

In the following derivations, we do not write dependencies of quantities on
$\CO{m}_O$ and $\din$ for clarity, indicating the relevant dependencies
at the end of each derivation.
We will also drop the explicit filtering of $J$ at $\kmax$, as we always only encounter the $\kmax$-filtered current.

\paragraph{Coefficients $\C$ in \refeq{Pexpand}.}

We start with
\ba
& \exp\bigg[ \sum_{m=3}^\infty \frac{1}{m!} \K^{\{m\},i_1\ldots i_m} J_{i_1} \cdots J_{i_m} \bigg] \vs
&= \sum_{N=0}^\infty \frac1{N!} \bigg[ \sum_{m=3}^\infty \frac{1}{m!} \K^{\{m\},i_1\ldots i_m} J_{i_1} \cdots J_{i_m} \bigg]^N\,.
\ea
In order to apply the multinomial formula, we restrict the sum over $m$ to $M$; we will specify $M$ once we reorder the sum below. Further, $N=0$ just yields the trivial 1 which we have pulled out in \refeq{Pexpand}, so we start with $N=1$. We have
\ba
& \sum_{N=1}^\infty \frac1{N!} \bigg[ \sum_{m=3}^{M} \frac{1}{m!} \K^{\{m\},i_1\ldots i_m} J_{i_1} \cdots J_{i_m} \bigg]^N \vs
&= \sum_{N=1}^\infty \sum_{\substack{a_3,\cdots,a_M \geq 0\\ a_3+\cdots + a_M=N}} 
\frac1{a_3! \cdots a_M!}
\prod_{m=3}^M
\bigg[ \frac{1}{m!} \K^{\{m\},i_1\ldots i_m} J_{i_1} \cdots J_{i_m} \bigg]^{a_m} \,.
\ea
Each term here has
\be
N_J = \sum_{m=3}^M m a_m =: L
\ee
powers of the current. We would now like to reorder the sum into an expansion in
powers of the current with coefficients $\C$,
\ba \label{eq:KtoC}
& \exp\bigg[ \sum_{m=3}^\infty \frac{1}{m!} \K^{\{m\},i_1\ldots i_m} J_{i_1} \cdots J_{i_m} \bigg] = \sum_{L=0}^\infty \frac1{L!} \C^{\{L\},i_1\ldots i_L} J_{i_1} \cdots J_{i_L}\,.
\ea
Abbreviating the indices (these can easily be restored, but become clumsy to write), we have for the coefficients
\ba
 \C^{\{L\},i_1\ldots i_L}
&= \sum_{N=1}^\infty \sum_{\substack{a_3,\cdots,a_M \geq 0\\ a_3+\cdots + a_M=N\\
3 a_3 + \cdots + M a_M = L} }
\frac{L!}{a_3! \cdots a_M!}
\prod_{m=3}^M
\left(\frac{1}{m!}\right)^{a_m}(\K^{\{m\} \cdots})^{a_m}\vs
&= \sum_{\substack{a_3,\cdots,a_L \geq 0\\ 3 a_3 + \cdots + L a_L = L} }
\frac{L!}{a_3! \cdots a_L!}
\prod_{m=3}^L
\left(\frac{1}{m!}\right)^{a_m}(\K^{\{m\} \cdots})^{a_m}
\,,
\ea
where $L \geq 3$ and we have used that the highest $M$ we need to consider is $M=L$, since the term with $a_L=1$ and $a_{m \neq L}=0$ already has the entire number of $L$ currents. This is \refeq{CCdef}. 

A few examples are given by 
\ba \label{eq:C_examples}
L=3:\qquad&
\C^{\{3\},i_1i_2i_3}[\{\CO{3}_O\},\din]
=  \K^{\{3\},i_1i_2i_3}[\{\CO{3}_O\},\din] \\
L=4:\qquad&
\C^{\{4\},i_1i_2i_3i_4}[\{\CO{4}_O\},\din]
=  \K^{\{4\},i_1i_2i_3i_4}[\{\CO{4}_O\},\din]  \vs
L=6:\qquad&
\C^{\{6\},i_1\ldots i_6}[\{\CO{3}_O\},\{\CO{6}_O\},\din]
=  \K^{\{6\},i_1\ldots i_6}[\{\CO{6}_O\},\din] \vs
&\hspace*{4.0cm}  + \frac{6!}{(3!)^2}  \K^{\{3\},i_1i_2i_3}[\{\CO{3}_O\},\din]
\K^{\{3\},i_4i_5i_6}[\{\CO{3}_O\},\din]\,.
\nonumber
\ea
In the following, we will also use the fully symmetrized version of the $\C^{\{L\}}$,
\be
\Csym^{\{L\},i_1\ldots i_L} \equiv  \frac1{L!} \sum_{\sigma\in S_L} \C^{\{L\},\sigma(i_1)\ldots \sigma(i_L)}\,,
\label{eq:Csym}
\ee
where $S_L$ is the group of permutations of $L$ elements (with cardinality $L!$). $\C^{\{3\}}$ and $\C^{\{4\}}$ are already symmetric thanks to the symmetry of the kernels $\K$, while for $L=6$, for example, we need to symmetrize by summing over the partitions of 6 indices into two groups of 3.

\paragraph{Coefficients $\tilde{\C}$ in \refeq{PexpandFormal}.}

We start from \refeq{PexpandInt},
\ba
\hatP[\hdgkmax] 
=  (Z[\v{J}=\v{0}])^{-1} \int& \Del\din \P[\din] \exp\bigg[-\frac12\v{\v{Y}^T \v{\Sigma}^{-1} \v{Y}}\bigg]
\int \Del (\ii \v{\tilde J}) 
\exp\bigg[\frac12 \v{\tilde J^T \Sigma \tilde J} \bigg] \vs
& \hspace{-3cm}\times\bigg[ 1 + \sum_{L=3}^\infty \frac{1}{L!} \C^{\{L\},i_1\ldots i_L}
  (\v{\tilde{J}} + \v{\Sigma}^{-1} \v{Y} )_{i_1} \cdots (\v{\tilde{J}} + \v{\Sigma}^{-1} \v{Y} )_{i_L} \bigg]\,.
\label{eq:appPprime0}
\ea
The main goal is to find $\tilde{\C}$, such that we can write it in the form of \refeq{PexpandFormal}, which we repeat here for convenience:
\ba
\hatP[\hdgkmax] 
= \N_{\LL} \int \Del\din \P[\din]& \exp\left[ -\frac12  \v{Y}^T \v{\Sigma}^{-1} \v{Y} \right]
\bigg[ 1 + \sum_{m=3}^\infty \frac1{m!} \tilde{\C}^{\{m\},i_1\ldots i_m}
Y_{i_1} \cdots Y_{i_m} \bigg]
\,.
\label{eq:PexpandFormalrepeat}
\ea
We first isolate the integral over $\v{\tilde{J}}$ for a single $L$. It is highly useful
to employ the symmetrized kernels $\Csym^{\{L\}}$, as we can then write the $\v{\tilde{J}}$ integrals in \refeq{appPprime0} as
\ba
\int \Del (\ii \v{\tilde J}) 
\exp\bigg[\frac12 \v{\tilde J^T \Sigma \tilde J} \bigg]
& \frac{1}{L!} \Csym^{\{L\},i_1\ldots i_L} \sum_{\ell=0}^L \binom{L}{\ell}
\tilde J_{i_1}\cdots \tilde J_{i_\ell}
  (\v{\Sigma}^{-1} \v{Y} )_{i_{\ell+1}} \cdots (\v{\Sigma}^{-1} \v{Y} )_{i_L}
\,.\nonumber
\ea
Now we perform the Gaussian integral over $\v{\tilde{J}}$, and drop a common
normalizing constant that is parameter independent, yielding
\ba
& |\v{\Sigma}|^{-1/2} \frac{1}{L!} \Csym^{\{L\},i_1\ldots i_L}  \sum_{\ell=0,\ \ell\ \rm even}^L \binom{L}{\ell} \frac{1}{2^{\ell/2} (\ell/2)!}
\sum_{\sigma\in S_\ell} \left(\v{\Sigma}^{-1}\right)_{\sigma(i_1)\sigma(i_2)} \cdots \left(\v{\Sigma}^{-1}\right)_{\sigma(i_{\ell-1})\sigma(i_\ell)} \vs
&\qquad\times (\v{\Sigma}^{-1} \v{Y} )_{i_{\ell+1}} \cdots (\v{\Sigma}^{-1} \v{Y} )_{i_L}
\,.
\ea
Given the symmetry of $\Csym^{\{L\}}$ in the $L$ indices, all permutations lead to the
same result, so we can cancel the factor of $(\ell/2)!$ to obtain, again for a fixed $L$, 
\ba
|\v{\Sigma}|^{-1/2} \frac1{L!}
 \sum_{\ell=0,\ \ell\ \rm even}^L &
\frac{L!}{2^{\ell/2} \ell!(L-\ell)!}
 \Csym^{\{L\},i_1\ldots i_L} \left(\v{\Sigma}^{-1}\right)_{i_1 i_2} \cdots \left(\v{\Sigma}^{-1}\right)_{i_{\ell-1} i_\ell} \vs
&\times (\v{\Sigma}^{-1} \v{Y} )_{i_{\ell+1}} \cdots (\v{\Sigma}^{-1} \v{Y} )_{i_L}
\,.
\ea
Noting that $|\v{\Sigma}|^{-1/2}$ is factored out into $\N_{\LL}$, and 
collecting all terms with $m$ powers of $Y$, and again using symmetry of the $\Csym^{\{L\}}$, we obtain for the coefficient
\ba
\tilde\C^{\{m\},i_1\ldots i_m} =  \sum_{\substack{L=m\\ L-m\ \rm even}}^\infty 
\frac{L!}{2^{(L-m)/2} (L-m)!}
 \Csym^{\{L\},j_1\ldots j_L}& \left(\v{\Sigma}^{-1}\right)_{j_{m+1} j_{m+2}} \cdots \left(\v{\Sigma}^{-1}\right)_{j_{L-1} j_{L}} \vs
&\times (\v{\Sigma}^{-1})_{j_1 i_1} \cdots (\v{\Sigma}^{-1})_{j_m i_m}
 \,,
 \label{eq:tildeCdef}
\ea
where recall that $\Csym^{\{L\}} = \Csym^{\{L\}}[\{\CO{\geq 3}\}, \din]$
while $\v{\Sigma} = \v{\Sigma}[\{\CO{2}_O\},\din]$, so that $\tilde\C^{\{m\}}=\tilde\C^{\{m\}}[\{\CO{\geq 2}\}, \din]$ (and $\v{Y} = \v{Y}[\hdgkmax, \{b_O\}, \din]$ as always).

At a given $m$, we can distinguish leading and higher-order contributions.
The leading order contribution is
\ba \label{eq:Ctilde_leading}
\tilde\C^{\{m\},i_1\ldots i_m} \stackrel{\rm LO}{=}  
 \Csym^{\{m\},j_1\ldots j_m}& (\v{\Sigma}^{-1})_{j_1 i_1} \cdots (\v{\Sigma}^{-1})_{j_m i_m}\,,
\ea
i.e.~it is given by $\Csym^{\{m\}}$ contracted with $m$ instances of $\v{\Sigma}^{-1}$. The next higher-order contribution is
\ba
\tilde\C^{\{m\},i_1\ldots i_m} \stackrel{\rm NLO}{=}  
\frac{(m+2)(m+1)}{2}
\Csym^{\{m+2\},j_1\ldots j_{m+2}}& \left(\v{\Sigma}^{-1}\right)_{j_{m+1} j_{m+2}}
(\v{\Sigma}^{-1})_{j_1 i_1} \cdots (\v{\Sigma}^{-1})_{j_m i_m}
\,,
 \label{eq:tildeCNLO}
\ea
consisting of $\Csym^{\{m+2\}}$ with two indices contracted with $\Sigma^{-1}$.
This similarly continues to higher order.

\section{Expansion of likelihood in the noise-field formulation}
\label{app:nsderivations}

Here we provide more explicit results and discussions on the posterior
expression in \refeq{nslike1m2}. Consider first the contribution from
$\ell=1$ to the second equality, dropping the normalization $\hat{\N}_{\LL}$ for convenience:
\ba
\hatP[\hdgkmax] \Big|_{\ell=1} &\propto -\int \Del\din \P[\din] \frac{\partial}{\partial Y^{k}}
\bigg\{ \sum_{m=2}^\infty \K^{\{m+1\},kj_1 \dots j_{m}}(\vKtwoinverse \v{Y})_{j_1} \dots (\vKtwoinverse \v{Y})_{j_{m}}
 \vs
 & \hspace{6cm}\times  
 \exp\left[ -\frac12 \v{Y}^T \vhatSigma^{-1}  \v{Y} \right] \bigg\} \vs
 &= -\int \Del\din \P[\din] \sum_{m=2}^\infty \K^{\{m+1\},kj_1 \dots j_{m}} \Biggr[ m
 (\vKtwoinverse \v{Y})_{j_1} \dots (\vKtwoinverse \v{Y})_{j_{m-1}} (\vKtwoinverse)_{j_m k} \vs
 & \hspace{1.5cm}
 - (\vKtwoinverse \v{Y})_{j_1} \dots (\vKtwoinverse \v{Y})_{j_m} (\vhatSigma^{-1} \v{Y})_{k} \Biggr]
 \exp\left[ -\frac12 \v{Y}^T \vhatSigma^{-1}  \v{Y} \right] \,.
 \label{eq:ell1}
\ea
Here we have used the symmetry of $\K^{\{m+1\},j_1\dots j_{m+1}}$. 
Comparing the contribution at $m+1$ with the definition of $\bb$,
\ba
\hatP[\hdgkmax] \propto \int \Del\din \P[\din] 
 \exp\left[ -\frac12  \v{Y}^T \vhatSigma^{-1} \v{Y} \right]
\left[1 + \sum_{m=1}^\infty \frac1{m!} \bb^{\{m\},i_1\ldots i_m} Y_{i_1}\ldots Y_{i_m}
  \right]\,,
 \ea
 we see that at $\ell=1$, we obtain contributions from $\K^{\{m+1\}}$ to $\bb^{\{m+1\}}$  [second term in \refeq{ell1}] and $\bb^{\{m-1\}}$ [first term in \refeq{ell1}].
Shifting $m$ by one, we can write
\ba 
\bb^{\{m\},j_1\dots j_{m}} \supset\:& m!\, \K^{\{m\},kj'_1 \dots j'_{m-1}} (\vKtwoinverse)_{j'_1 j_1} \dots (\vKtwoinverse)_{j'_{m-1} j_{m-1}} (\vhatSigma^{-1})_{kj_{m}} \,,\vs
\bb^{\{m-2\},j_1\dots j_{m-2}} \supset\:&  -(m-2)! (m-1) \, \K^{\{m\},kj'_1 \dots j'_{m-1}} (\vKtwoinverse)_{j_1'j_1} \cdots (\vKtwoinverse)_{j'_{m-2}j_{m-2}} \vs
&\hspace*{4cm} \times (\vKtwoinverse)_{j'_{m-1} k} \,,\label{eq:bbexamples_app}
\ea
where $\vKtwoinverse = \vKtwoinverse[\{\beps{1}_O\}, \din]$. 
The first line here provides a one-to-one mapping between $\K^{\{m\}}$ and
$\bb^{\{m\}}$.\footnote{The additional $m!$ factor appearing here can be understood by looking at the stochastic contribution to the $(m+1)$-point function, which is proportional to $m! \beps{m}_\unit$ in the noise-field formulation (cf. \refapp{npt}), whereas it is proportional to $\CO{m+1}_\unit$ in the partition function description [cf. \refeq{dd_general}]. One could alternatively define the $\beps{m}_O$ with an additional $1/m!$.} The contractions with $\vKtwoinverse$ are analogous to the
contractions with $\Sigma^{-1}$ 
appearing in the expansion around the Gaussian likelihood [cf. the definition
of the $\tilde\C^{\{m\}}$, \refeq{tildeCdef}]. 
In addition, there are contributions to $\bb^{\{m-2\}}$. These correspond to
a shift of lower-order contributions by higher-order ones, analogous
to \refeq{tildeCNLO}.
Notice that, for the lowest non-Gaussian stochastic term
with $m=3$, we obtain a unique contribution to $\bb^{\{1\}}$:
\be
\bb^{\{1\},j} = -2 \, \K^{\{3\},kj' j'_2} (\vKtwoinverse)_{j'j} (\vKtwoinverse)_{j'_{2} k} \,,
\ee
where $\K^{\{3\}} = \K^{\{3\}}[\{\beps{2}_O\},\din]$. 
As mentioned in \refsec{NS}, this contribution can be removed by redefining
\be \label{eq:Yremap} 
Y^l \to Y^l - 2 (\vKtwoinverse)^{lj}(\vKtwoinverse)_{jj'} \K^{\{3\},kj' j'_2}  (\vKtwoinverse)_{j'_{2} k}\,,
\ee
where the shifted $\v{Y}$ is now a function of $\{\beps{1}_O,\beps{2}_O\}$
in addition to $\{\beps{0}_O\}$. Similarly, the term $\bb^{\{2\}}$ can be absorbed by a shift in $\vhatSigma$.

Finally, we turn to $\ell>1$. First, notice that contributions at order $\ell$
involve $\ell$ powers of the kernels $\K^{\{m\geq 3\}}$, so that they will provide
subleading corrections to the relation between the $\K^{\{m\}}$ and $\bb^{\{m\}}$.
Further, it is straightforward to see that, due to the same additional factors
of $\K^{\{m\geq 3\}} Y^{m-1}$ that come in at each $\ell$, the lowest polynomial
order in $Y$ that can be reached at a given $\ell$ is $\ell$. That is,
in order to derive the expression for the coefficient $\bb^{\{m\}}$ it is sufficient to
consider $1\leq \ell \leq m$.

\bibliographystyle{JHEP}
\bibliography{main}

\providecommand{\href}[2]{#2}\begingroup\raggedright\begin{thebibliography}{10}

\bibitem{Desjacques:2016bnm}
V.~Desjacques, D.~Jeong and F.~Schmidt, \emph{{Large-Scale Galaxy Bias}},
  \href{http://dx.doi.org/10.1016/j.physrep.2017.12.002}{\emph{Phys. Rept.}
  {\bf 733} (2018) 1--193}, [\href{http://arxiv.org/abs/1611.09787}{{\tt
  1611.09787}}].

\bibitem{MSZ}
M.~{Mirbabayi}, F.~{Schmidt} and M.~{Zaldarriaga}, \emph{{Biased tracers and
  time evolution}},
  \href{http://dx.doi.org/10.1088/1475-7516/2015/07/030}{\emph{\jcap} {\bf 7}
  (July, 2015) 30}, [\href{http://arxiv.org/abs/1412.5169}{{\tt 1412.5169}}].

\bibitem{DAmico:2022ukl}
G.~D'Amico, Y.~Donath, M.~Lewandowski, L.~Senatore and P.~Zhang, \emph{{The
  one-loop bispectrum of galaxies in redshift space from the Effective Field
  Theory of Large-Scale Structure}},
  \href{http://arxiv.org/abs/2211.17130}{{\tt 2211.17130}}.

\bibitem{Rubira:2024tea}
H.~Rubira and F.~Schmidt, \emph{{The renormalization group for large-scale
  structure: origin of galaxy stochasticity}},
  \href{http://dx.doi.org/10.1088/1475-7516/2024/10/092}{\emph{JCAP} {\bf 10}
  (2024) 092}, [\href{http://arxiv.org/abs/2404.16929}{{\tt 2404.16929}}].

\bibitem{Hamaus:2010im}
N.~Hamaus, U.~Seljak, V.~Desjacques, R.~E. Smith and T.~Baldauf,
  \emph{{Minimizing the Stochasticity of Halos in Large-Scale Structure
  Surveys}}, \href{http://dx.doi.org/10.1103/PhysRevD.82.043515}{\emph{Phys.
  Rev. D} {\bf 82} (2010) 043515}, [\href{http://arxiv.org/abs/1004.5377}{{\tt
  1004.5377}}].

\bibitem{Rubira:2025scu}
H.~Rubira and F.~Conteddu, \emph{{Multi-tracer beyond linear theory}},
  \href{http://dx.doi.org/10.1088/1475-7516/2025/10/111}{\emph{JCAP} {\bf 10}
  (2025) 111}, [\href{http://arxiv.org/abs/2504.18245}{{\tt 2504.18245}}].

\bibitem{Voivodic:2025quw}
R.~Voivodic, \emph{{Perturbative Likelihoods for Large-Scale Structure of the
  Universe}},  \href{http://arxiv.org/abs/2505.23750}{{\tt 2505.23750}}.

\bibitem{2013MNRAS.432..894J}
J.~{Jasche} and B.~D. {Wandelt}, \emph{{Bayesian physical reconstruction of
  initial conditions from large-scale structure surveys}},
  \href{http://dx.doi.org/10.1093/mnras/stt449}{\emph{Mon. Not. Roy. Astron.
  Soc.} {\bf 432} (June, 2013) 894--913},
  [\href{http://arxiv.org/abs/1203.3639}{{\tt 1203.3639}}].

\bibitem{2013MNRAS.429L..84K}
F.~S. {Kitaura}, \emph{{The initial conditions of the universe from constrained
  simulations.}}, \href{http://dx.doi.org/10.1093/mnrasl/sls029}{\emph{Mon.
  Not. Roy. Astron. Soc.} {\bf 429} (Feb., 2013) L84--L88},
  [\href{http://arxiv.org/abs/1203.4184}{{\tt 1203.4184}}].

\bibitem{2013ApJ...772...63W}
H.~{Wang}, H.~J. {Mo}, X.~{Yang} and F.~C. {van den Bosch},
  \emph{{Reconstructing the Initial Density Field of the Local Universe:
  Methods and Tests with Mock Catalogs}},
  \href{http://dx.doi.org/10.1088/0004-637X/772/1/63}{\emph{Astrophys. J.} {\bf
  772} (July, 2013) 63}, [\href{http://arxiv.org/abs/1301.1348}{{\tt
  1301.1348}}].

\bibitem{Wang:2014hia}
H.~Wang, H.~J. Mo, X.~Yang, Y.~P. Jing and W.~P. Lin, \emph{{ELUCID - Exploring
  the Local Universe with reConstructed Initial Density field I: Hamiltonian
  Markov Chain Monte Carlo Method with Particle Mesh Dynamics}},
  \href{http://dx.doi.org/10.1088/0004-637X/794/1/94}{\emph{Astrophys. J.} {\bf
  794} (2014) 94}, [\href{http://arxiv.org/abs/1407.3451}{{\tt 1407.3451}}].

\bibitem{Modi:2018cfi}
C.~Modi, Y.~Feng and U.~Seljak, \emph{{Cosmological Reconstruction From Galaxy
  Light: Neural Network Based Light-Matter Connection}},
  \href{http://dx.doi.org/10.1088/1475-7516/2018/10/028}{\emph{JCAP} {\bf 10}
  (2018) 028}, [\href{http://arxiv.org/abs/1805.02247}{{\tt 1805.02247}}].

\bibitem{Schmidt:2020viy}
F.~Schmidt, G.~Cabass, J.~Jasche and G.~Lavaux, \emph{{Unbiased Cosmology
  Inference from Biased Tracers using the EFT Likelihood}},
  \href{http://dx.doi.org/10.1088/1475-7516/2020/11/008}{\emph{JCAP} {\bf 11}
  (2020) 008}, [\href{http://arxiv.org/abs/2004.06707}{{\tt 2004.06707}}].

\bibitem{Schmidt:2020tao}
F.~Schmidt, \emph{{Sigma-Eight at the Percent Level: The EFT Likelihood in Real
  Space}}, \href{http://dx.doi.org/10.1088/1475-7516/2021/04/032}{\emph{JCAP}
  {\bf 04} (2021) 032}, [\href{http://arxiv.org/abs/2009.14176}{{\tt
  2009.14176}}].

\bibitem{Shallue:2022mhf}
C.~J. Shallue and D.~J. Eisenstein, \emph{{Reconstructing cosmological initial
  conditions from late-time structure with convolutional neural networks}},
  \href{http://dx.doi.org/10.1093/mnras/stad528}{\emph{Mon. Not. Roy. Astron.
  Soc.} {\bf 520} (2023) 6256--6267},
  [\href{http://arxiv.org/abs/2207.12511}{{\tt 2207.12511}}].

\bibitem{Chen:2023uup}
X.~Chen, F.~Zhu, S.~Gaines and N.~Padmanabhan, \emph{{Effective cosmic density
  field reconstruction with convolutional neural network}},
  \href{http://dx.doi.org/10.1093/mnras/stad1868}{\emph{Mon. Not. Roy. Astron.
  Soc.} {\bf 523} (2023) 6272--6281},
  [\href{http://arxiv.org/abs/2306.10538}{{\tt 2306.10538}}].

\bibitem{Andrews:2022nvv}
A.~Andrews, J.~Jasche, G.~Lavaux and F.~Schmidt, \emph{{Bayesian field-level
  inference of primordial non-Gaussianity using next-generation galaxy
  surveys}}, \href{http://dx.doi.org/10.1093/mnras/stad432}{\emph{Mon. Not.
  Roy. Astron. Soc.} {\bf 520} (2023) 5746--5763},
  [\href{http://arxiv.org/abs/2203.08838}{{\tt 2203.08838}}].

\bibitem{Modi:2022pzm}
C.~Modi, Y.~Li and D.~Blei, \emph{{Reconstructing the universe with variational
  self-boosted sampling}},
  \href{http://dx.doi.org/10.1088/1475-7516/2023/03/059}{\emph{JCAP} {\bf 03}
  (2023) 059}, [\href{http://arxiv.org/abs/2206.15433}{{\tt 2206.15433}}].

\bibitem{Dai:2022dso}
B.~Dai and U.~Seljak, \emph{{Translation and rotation equivariant normalizing
  flow (TRENF) for optimal cosmological analysis}},
  \href{http://dx.doi.org/10.1093/mnras/stac2010}{\emph{Mon. Not. Roy. Astron.
  Soc.} {\bf 516} (2022) 2363--2373},
  [\href{http://arxiv.org/abs/2202.05282}{{\tt 2202.05282}}].

\bibitem{Kostic:2022vok}
A.~Kosti{\'c}, N.-M. Nguyen, F.~Schmidt and M.~Reinecke, \emph{{Consistency
  tests of field level inference with the EFT likelihood}},
  \href{http://dx.doi.org/10.1088/1475-7516/2023/07/063}{\emph{JCAP} {\bf 07}
  (2023) 063}, [\href{http://arxiv.org/abs/2212.07875}{{\tt 2212.07875}}].

\bibitem{Qin:2023dew}
F.~Qin, D.~Parkinson, S.~E. Hong and C.~G. Sabiu, \emph{{Reconstructing the
  cosmological density and velocity fields from redshifted galaxy distributions
  using V-net}},
  \href{http://dx.doi.org/10.1088/1475-7516/2023/06/062}{\emph{JCAP} {\bf 06}
  (2023) 062}, [\href{http://arxiv.org/abs/2302.02087}{{\tt 2302.02087}}].

\bibitem{Jindal:2023qew}
V.~Jindal, D.~Jamieson, A.~Liang, A.~Singh and S.~Ho, \emph{{Predicting the
  Initial Conditions of the Universe using Deep Learning}},
  \href{http://arxiv.org/abs/2303.13056}{{\tt 2303.13056}}.

\bibitem{Charnock:2019rbk}
T.~Charnock, G.~Lavaux, B.~D. Wandelt, S.~Sarma~Boruah, J.~Jasche and M.~J.
  Hudson, \emph{{Neural physical engines for inferring the halo mass
  distribution function}},
  \href{http://dx.doi.org/10.1093/mnras/staa682}{\emph{Mon. Not. Roy. Astron.
  Soc.} {\bf 494} (2020) 50--61}, [\href{http://arxiv.org/abs/1909.06379}{{\tt
  1909.06379}}].

\bibitem{Doeser:2023yzv}
L.~Doeser, D.~Jamieson, S.~Stopyra, G.~Lavaux, F.~Leclercq and J.~Jasche,
  \emph{{Bayesian inference of initial conditions from non-linear cosmic
  structures using field-level emulators}},
  \href{http://dx.doi.org/10.1093/mnras/stae2429}{\emph{Mon. Not. Roy. Astron.
  Soc.} {\bf 535} (2024) 1258--1277},
  [\href{http://arxiv.org/abs/2312.09271}{{\tt 2312.09271}}].

\bibitem{Babic:2024wph}
I.~Babi{\'c}, F.~Schmidt and B.~Tucci, \emph{{Straightening the Ruler:
  Field-Level Inference of the BAO Scale with LEFTfield}},
  \href{http://arxiv.org/abs/2407.01524}{{\tt 2407.01524}}.

\bibitem{FBISBI}
N.-M. {Nguyen}, F.~{Schmidt}, B.~{Tucci}, M.~{Reinecke} and A.~{Kosti{\'c}},
  \emph{{How Much Information Can Be Extracted from Galaxy Clustering at the
  Field Level?}},
  \href{http://dx.doi.org/10.1103/PhysRevLett.133.221006}{\emph{\prl} {\bf 133}
  (Nov., 2024) 221006}, [\href{http://arxiv.org/abs/2403.03220}{{\tt
  2403.03220}}].

\bibitem{Cabass:2019lqx}
G.~Cabass and F.~Schmidt, \emph{{The EFT Likelihood for Large-Scale
  Structure}},
  \href{http://dx.doi.org/10.1088/1475-7516/2020/04/042}{\emph{JCAP} {\bf 04}
  (2020) 042}, [\href{http://arxiv.org/abs/1909.04022}{{\tt 1909.04022}}].

\bibitem{Cabass:2020nwf}
G.~Cabass and F.~Schmidt, \emph{{The Likelihood for LSS: Stochasticity of Bias
  Coefficients at All Orders}},
  \href{http://dx.doi.org/10.1088/1475-7516/2020/07/051}{\emph{JCAP} {\bf 07}
  (2020) 051}, [\href{http://arxiv.org/abs/2004.00617}{{\tt 2004.00617}}].

\bibitem{Rubira:2023vzw}
H.~Rubira and F.~Schmidt, \emph{{Galaxy bias renormalization group}},
  \href{http://dx.doi.org/10.1088/1475-7516/2024/01/031}{\emph{JCAP} {\bf 01}
  (2024) 031}, [\href{http://arxiv.org/abs/2307.15031}{{\tt 2307.15031}}].

\bibitem{Nikolis:2024kbx}
C.~Nikolis, H.~Rubira and F.~Schmidt, \emph{{The renormalization group for
  large-scale structure: primordial non-Gaussianities}},
  \href{http://dx.doi.org/10.1088/1475-7516/2024/08/017}{\emph{JCAP} {\bf 08}
  (2024) 017}, [\href{http://arxiv.org/abs/2405.21002}{{\tt 2405.21002}}].

\bibitem{Bakx:2025cvu}
T.~Bakx, M.~Garny, H.~Rubira and Z.~Vlah, \emph{{Two-loop renormalization and
  running of galaxy bias}},  \href{http://arxiv.org/abs/2507.13905}{{\tt
  2507.13905}}.

\bibitem{Tucci:2023bag}
B.~{Tucci} and F.~{Schmidt}, \emph{{EFTofLSS meets simulation-based inference:
  {\ensuremath{\sigma}} $_{8}$ from biased tracers}},
  \href{http://dx.doi.org/10.1088/1475-7516/2024/05/063}{\emph{\jcap} {\bf
  2024} (May, 2024) 063}, [\href{http://arxiv.org/abs/2310.03741}{{\tt
  2310.03741}}].

\bibitem{Schmidt:2025iwa}
F.~Schmidt, \emph{{On the connection between field-level inference and n-point
  correlation functions}},
  \href{http://dx.doi.org/10.1088/1475-7516/2025/09/056}{\emph{JCAP} {\bf 09}
  (2025) 056}, [\href{http://arxiv.org/abs/2504.15351}{{\tt 2504.15351}}].

\bibitem{Schmidt:2018bkr}
F.~Schmidt, F.~Elsner, J.~Jasche, N.~M. Nguyen and G.~Lavaux, \emph{{A rigorous
  EFT-based forward model for large-scale structure}},
  \href{http://dx.doi.org/10.1088/1475-7516/2019/01/042}{\emph{JCAP} {\bf 01}
  (2019) 042}, [\href{http://arxiv.org/abs/1808.02002}{{\tt 1808.02002}}].

\bibitem{Schmittfull:2018yuk}
M.~Schmittfull, M.~Simonovi\'c, V.~Assassi and M.~Zaldarriaga, \emph{{Modeling
  Biased Tracers at the Field Level}},
  \href{http://dx.doi.org/10.1103/PhysRevD.100.043514}{\emph{Phys. Rev. D} {\bf
  100} (2019) 043514}, [\href{http://arxiv.org/abs/1811.10640}{{\tt
  1811.10640}}].

\bibitem{akitsu/etal:2025}
K.~Akitsu, M.~Simonovi{\'c}, S.-F. Chen, G.~Cabass and M.~Zaldarriaga,
  \emph{{Cosmology inference with perturbative forward modeling at the field
  level: a comparison with joint power spectrum and bispectrum analyses}},
  \href{http://arxiv.org/abs/2509.09673}{{\tt 2509.09673}}.

\bibitem{beyond2pt}
{\scshape Beyond-2pt} collaboration, E.~Krause et~al., \emph{{A
  Parameter-Masked Mock Data Challenge for Beyond-Two-Point Galaxy Clustering
  Statistics}},
  \href{http://dx.doi.org/10.3847/1538-4357/ad781d}{\emph{Astrophys. J.} {\bf
  990} (2025) 99}, [\href{http://arxiv.org/abs/2405.02252}{{\tt 2405.02252}}].

\bibitem{elsner/etal}
F.~{Elsner}, F.~{Schmidt}, J.~{Jasche}, G.~{Lavaux} and N.-M. {Nguyen},
  \emph{{Cosmology inference from a biased density field using the EFT-based
  likelihood}},
  \href{http://dx.doi.org/10.1088/1475-7516/2020/01/029}{\emph{\jcap} {\bf
  2020} (Jan., 2020) 029}, [\href{http://arxiv.org/abs/1906.07143}{{\tt
  1906.07143}}].

\bibitem{stadler:2024a}
J.~Stadler, F.~Schmidt and M.~Reinecke, \emph{{Fast, accurate and perturbative
  forward modeling of galaxy clustering. Part I. Galaxies in the restframe}},
  \href{http://dx.doi.org/10.1088/1475-7516/2025/04/089}{\emph{JCAP} {\bf 04}
  (2025) 089}, [\href{http://arxiv.org/abs/2409.10937}{{\tt 2409.10937}}].

\bibitem{Cabass:2020jqo}
G.~Cabass, \emph{{The EFT Likelihood for Large-Scale Structure in Redshift
  Space}}, \href{http://dx.doi.org/10.1088/1475-7516/2021/01/067}{\emph{JCAP}
  {\bf 01} (2021) 067}, [\href{http://arxiv.org/abs/2007.14988}{{\tt
  2007.14988}}].

\bibitem{stadler:2024b}
J.~Stadler, F.~Schmidt, M.~Reinecke and M.~Esposito, \emph{{Fast, Accurate and
  Perturbative Forward Modeling of Galaxy Clustering Part II: Redshift Space}},
   \href{http://arxiv.org/abs/2411.04513}{{\tt 2411.04513}}.

\bibitem{nguyen/etal:2020}
N.-M. {Nguyen}, F.~{Schmidt}, G.~{Lavaux} and J.~{Jasche}, \emph{{Impacts of
  the physical data model on the forward inference of initial conditions from
  biased tracers}},
  \href{http://dx.doi.org/10.1088/1475-7516/2021/03/058}{\emph{\jcap} {\bf
  2021} (Mar., 2021) 058}, [\href{http://arxiv.org/abs/2011.06587}{{\tt
  2011.06587}}].

\bibitem{Baldauf:2013hka}
T.~Baldauf, U.~Seljak, R.~E. Smith, N.~Hamaus and V.~Desjacques, \emph{{Halo
  stochasticity from exclusion and nonlinear clustering}},
  \href{http://dx.doi.org/10.1103/PhysRevD.88.083507}{\emph{Phys. Rev. D} {\bf
  88} (2013) 083507}, [\href{http://arxiv.org/abs/1305.2917}{{\tt 1305.2917}}].

\end{thebibliography}\endgroup

\end{document}